\documentclass[fleqn,usenatbib]{mnras}

\usepackage{newtxtext,newtxmath}

\usepackage[T1]{fontenc}
\usepackage{ae,aecompl}

\usepackage{graphicx}
\usepackage{amsmath}
\graphicspath{{./Images/}}

\newcommand{\msun}{M$_{\odot}$}
\newcommand{\rsun}{R$_{\odot}$}
\usepackage{deluxetable}
\usepackage{color}
\definecolor{cerulean}{rgb}{0.0, 0.48, 0.65}

\makeatletter
\def\@to{to}
\makeatother

\title[Aligning Orbits in AGN]{Aligning Nuclear Cluster Orbits with an Active Galactic Nucleus Accretion Disk}

\author[G. Fabj et al.]{
Gaia Fabj,$^{1,2,3,4}$\thanks{E-mail:  gaia.fabj@stud.uni-heidelberg.de}
Syeda S. Nasim,$^{4,5}$\thanks{E-mail:  snasim@amnh.org}
Freddy Caban,$^{6,4}$
K. E. Saavik Ford,$^{7,4,9}$
\newauthor Barry McKernan,$^{7,4,9}$ and Jillian M. Bellovary$^{8,4,9}$
\\
$^{1}$Astronomisches Rechen-Institut, Zentrum f\"{u}r Astronomie, Universit\"{a}t Heidelberg, 69120 Heidelberg, Germany \\
$^{2}$Dept. of Physics and Astronomy, Universit\"{a}t Heidelberg, 69117 Heidelberg, Germany\\
$^{3}$Dept. of Physics and Astronomy, College of Staten Island, City University of New York, Staten Island, NY 10314 USA\\
$^{4}$Dept. of Astrophysics, American Museum of Natural History, New York, NY 10024 USA\\
$^{5}$Dept. of Physics and Astronomy, Hunter College, City University of New York, New York, NY 10065 USA\\
$^{6}$Dept. of Physics, Queens College, City University of New York, Flushing, NY 11367 USA\\
$^{7}$Dept. of Science, Borough of Manhattan Community College, City University of New York, New York, NY 10007 USA\\
$^{8}$Dept. of Physics, Queensborough Community College, City University of New York, Bayside, NY 11364 USA\\
$^{9}$Physics Program, CUNY Graduate Center, City University of New York, New York, NY 10016 USA}

\date{Accepted 2020 September 23. Received  2020 September 6; in original form 2020 June 22}

\pubyear{2020}
\volume{499}
\pagerange{2608-2616}

\begin{document}
\label{firstpage}
\pagerange{\pageref{firstpage}--\pageref{lastpage}}
\maketitle

\begin{abstract}
Active galactic nuclei (AGN) are powered by the accretion of disks of gas onto supermassive black holes (SMBHs). Stars and stellar remnants orbiting the SMBH in the nuclear star cluster (NSC) will interact with the AGN disk. Orbiters plunging through the disk experience a drag force and, through repeated passage, can have their orbits captured by the disk.  A population of embedded objects in AGN disks may be a significant source of binary black hole mergers, supernovae, tidal disruption events and embedded gamma-ray bursts. For two representative AGN disk models we use geometric drag and Bondi-Hoyle-Littleton drag to determine the time to capture for stars and stellar remnants. We assume a range of initial inclination angles and semi-major axes for circular Keplerian prograde orbiters. Capture time strongly depends on the density and aspect ratio of the chosen disk model, the relative velocity of the stellar object with respect to the disk, and the AGN lifetime. We expect that for an AGN disk density $\rho \gtrsim 10^{-11}{\rm g\,cm^{-3}}$ and disk lifetime $\geq 1$Myr, there is a significant population of embedded stellar objects, which can fuel mergers detectable in gravitational waves with LIGO-Virgo and LISA.
\end{abstract}

\begin{keywords}

stars: kinematics and dynamics -- galaxies: active -- accretion, accretion discs -- galaxies: nuclei -- stars: black holes -- gravitational waves

\end{keywords}

\section{Introduction} \label{sec:introduction}

Galactic nuclei contain large numbers of stars and stellar remnants orbiting the central supermassive black hole (SMBH) \citep[e.g.][]{Morris93, MiraldaEscude00, Hailey18, Generozov18}. SMBH orbiters are dynamically `hot', with a large velocity dispersion \citep{Antonini14,Leigh18}. However, a small fraction of galactic nuclei are active due to a disky gas accretion flow onto the SMBH. The resulting active galactic nucleus (AGN) disk, acts to dynamically `cool' SMBH orbiters over its lifetime, generating an orbiter velocity dispersion that approaches that of the gas disk \citep{Ostriker83, Syer91, McKernan12}. 

AGN disks therefore contain a two-component population of embedded objects: an initial component corresponding to orbiters that geometrically coincide with the disk, and a captured component that grows over time as orbits are dragged into alignment with the disk plane \citep[e.g.][]{Syer91,Artymowicz93,Rauch95,McKernan12,Just12,Kennedy16,Bekdaulet2018,MacLeod20}. The initial embedded population is simply a function of the geometric size and aspect ratio of the disk. The captured component will grow as a function of the drag experienced by disk-crossing orbiters, which in turn depends on the nuclear population, the disk density profile, geometric thickness and lifetime.

Objects embedded in AGN (or protoplanetary) disks will experience torques from the disk gas and migrate \citep{LinPap86, Ward_1997,Tanaka_2002,Levin_2007, Paardekooper_2010, Lyra_2010, Horn_2012}. Migrator encounters promote the formation of new binaries \citep{Secunda19,Secunda20} and gas hardening can drive rapid mergers \citep{Baruteau11}, yielding gravitational wave (GW) events detectable with LIGO-Virgo \citep[e.g.][]{McKernan12, McKernan14, Bellovary16,Bartos17, Stone17, McKernan18, McK19a, Yang19,McK20,Tagawa20,Grobner20,Ishibashi20} and with LISA \citep{McKernan14,Derdzinski19,Derdzinski20}. By disentangling the AGN contribution to LIGO-Virgo detection rates, it will be possible to probe beneath the AGN photosphere and place strong constraints on AGN mid-plane gas densities, disk aspect ratios and disk lifetimes.  

If AGN disks are efficient at nuclear orbit capture, the  contribution of this channel to GW events detectable with LIGO-Virgo and LISA is likely to be significant and possibly dominant.Efficient disk capture is important for the growth of hierarchical disk mergers at migration traps \citep{McK20,Secunda20}, a possible environment for recent merger detection GW190521 \citep{GWa,GWb}. Efficient disk capture also increases the number of electromagnetic (EM) transients associated with embedded objects in AGN, that might be detectable in large sky surveys of AGNs \citep[e.g.][]{Graham17,Cannizzaro20}. The process of orbit capture may also lead to detectable EM signatures. Grazing inclination orbits encountering the disk may generate luminous counterparts from Bondi drag shocks \citep{Graham20b}. 

Here we investigate the rate of capture of stars and stellar-origin black holes (sBH) into AGN disks for two representative disk models. While multiple mechanisms may be responsible for the dynamical cooling of orbits, we focus here on the impact of drag, in particular Geometric drag for stars and Bondi-Hoyle-Lyttleton drag for sBH. This paper is laid out as follows. In \S\ref{sec:methods} we describe the models and methods underlying our calculations. In \S\ref{sec:results} we describe our results, and in \S\ref{sec:discussion} we discuss our results in the context of other work along with the implications for gravitational wave observations. Finally, in \S\ref{sec:conclusions} we outline our conclusions.

\section{Models and Methods} \label{sec:methods}

Here we outline our assumptions concerning nuclear star clusters, AGN disk models and drag forces at work for disk-crossing orbiters. 

\subsection{Nuclear Star Clusters} \label{sec:nsc}

Since we are concerned with the alignment of orbiters in galactic nuclei with AGN disks, we must first consider properties of nuclear star clusters (NSCs). NSCs are found in many (particularly dwarf) galactic nuclei in the local Universe \citep[e.g.][]{Boker02,Cote06,Wehner06}. NSCs and SMBHs co-exist, with NSCs becoming less significant or absent for $M_{\rm SMBH}>5 \times 10^{7}M_{\odot}$ \citep{Graham09}. The $M-\sigma$ relation may be different for NSCs than for SMBH \citep[e.g.][]{McLaughlin06,Leigh12,Scott13}.
 
NSCs are significantly brighter and denser than globular clusters, but have the same size scale \citep[e.g.][and references therein]{Neumayer20}. In our own Galactic nucleus, the NSC may have been built up from the infall of multiple globular clusters plus nuclear star formation, over a Hubble time \citep[e.g.][]{Antonini14}. Two-body interactions and the tendency of the system to approach equipartition causes massive objects to sink and less massive objects to diffuse outwards \citep{BahcallWolf76}, causing mass segregation. For a low initial density of larger mass objects, the less massive objects in a galactic nucleus act as a dynamical friction background and a steep mass segregation profile can develop, with a very high number density of stellar origin black holes (sBH) in the central regions \citep[e.g.][]{Alexander09,PretoAmaro10}.  

\begin{deluxetable}{lcc}
\tablecolumns{3} 
\tablewidth{170pt}
\tablecaption{Assumed Properties of NSC Orbiters}
\tablehead{ \colhead{Object} & \colhead{Mass [\msun]} & \colhead{Radius [\rsun]} }
\startdata 
M Dwarf & 0.5 & 0.4\\
G Star & 1 & 1\\
O Star & 50 & 15\\
Red Giant & 1.5 & 100\\
sBH & 10 & -\\
\enddata
\label{table:properties}
\end{deluxetable}

Here we consider the interaction between NSC elements and an AGN disk with a central $10^8 M_{\odot}$ SMBH. \footnote{For consistency and comparison with SG we rescale TQM to a $10^8M_{\odot}$ SMBH even though the model is originally constructed for a central object of $10^9 M_{\odot}$.}
We assume that there is a dense population of stars (of various types) and stellar remnants (white dwarfs, neutron stars and sBH) that interacts with our model AGN disks. Table~\ref{table:properties} shows the properties of fiducial NSC orbiters.

In the absence of an AGN disk, the orbits will relax into a dynamically hot state with a distribution of orbital parameters: semi-major axes ($a$), eccentricity ($e$), and inclination angle ($i$) \citep{Antonini14}. In this work, we shall assume Keplerian orbits spanning a range of ($a,i$) for all stellar types and sBH and allow these to evolve due to disk drag effects (i.e. $\frac {da}{dt}, \frac {di}{dt} <0$). For simplicity, we shall assume $e=0$ and $\frac {de}{dt}=0$ for all orbits. See Section \ref{sec:comparison} for a discussion of the consequences of our assumptions for $(e,de/dt)$ \citep[and see also][]{MacLeod20}.

\subsection{AGN disk models} \label{sec:models}

\begin{figure}
\includegraphics[width=\linewidth]{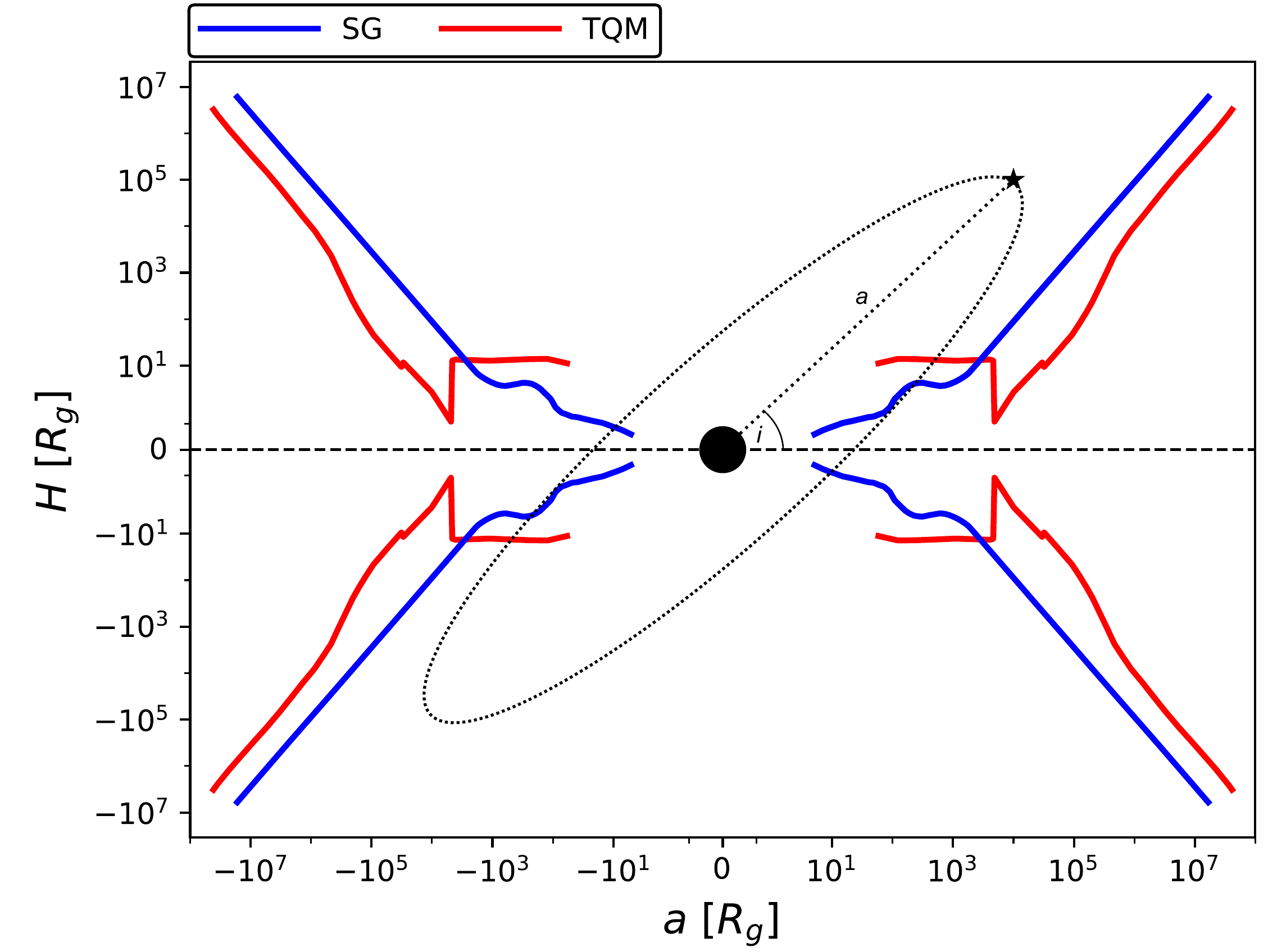}
\includegraphics[width=\linewidth]{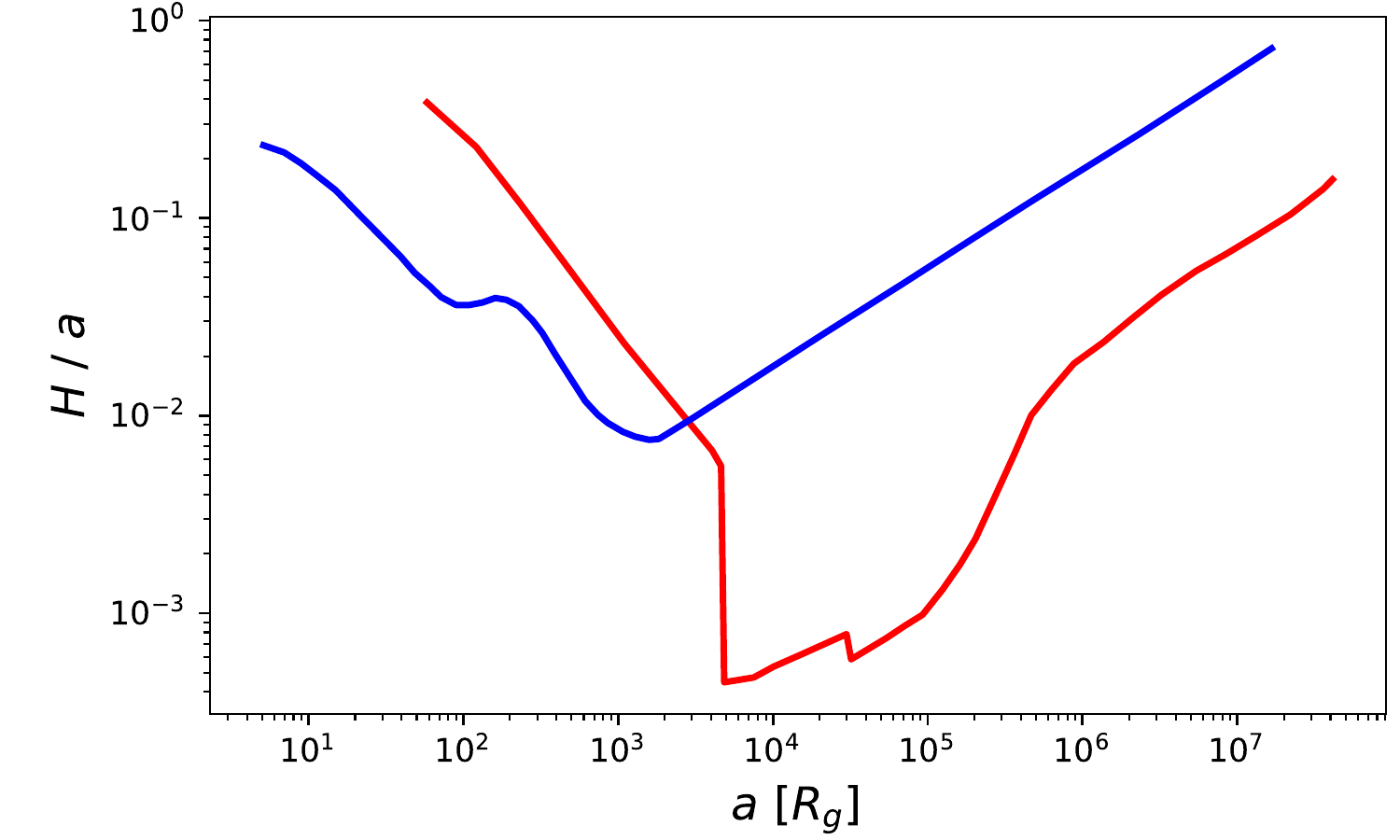}
\includegraphics[width=\linewidth]{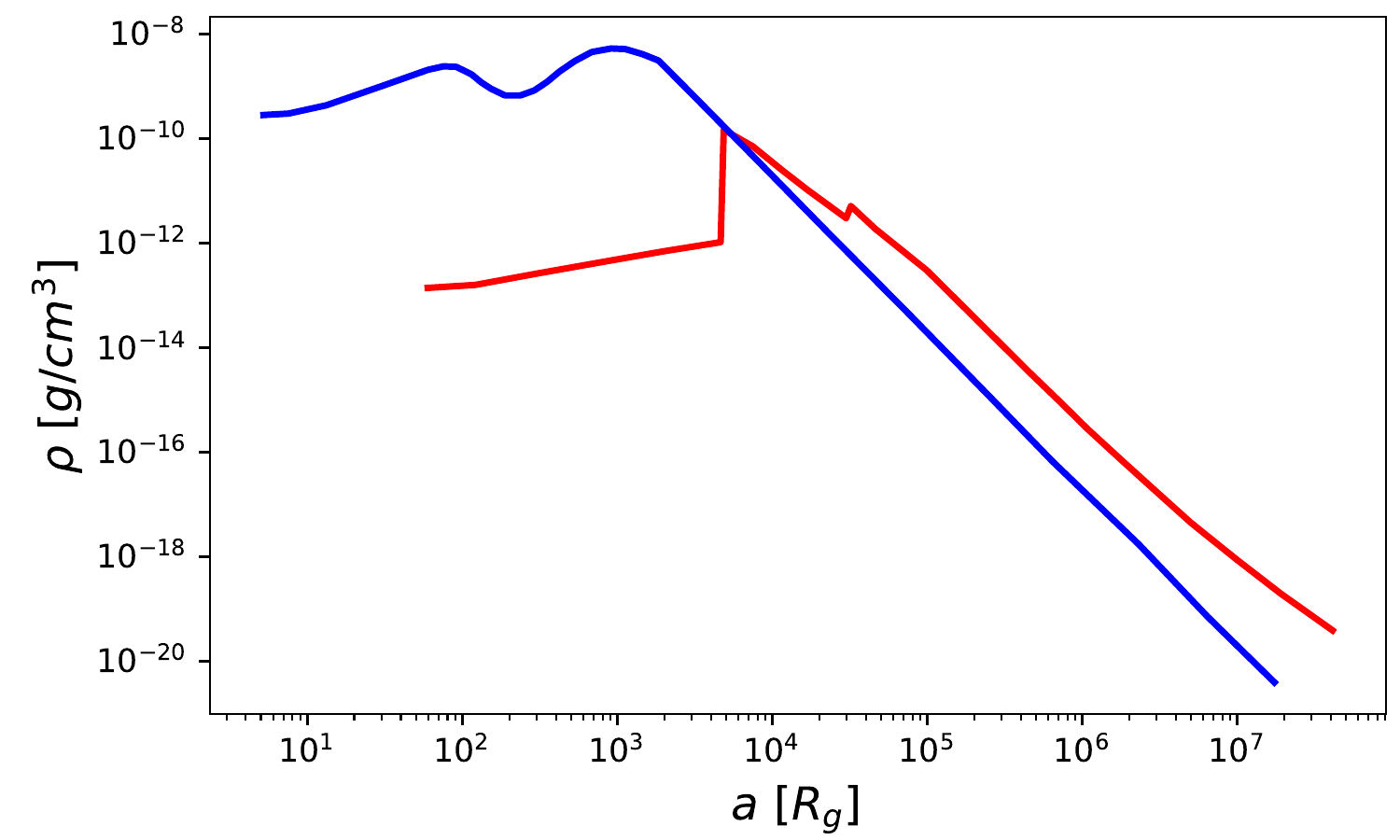}
\caption{Disk height (top), aspect ratio (middle), and density (bottom) profiles as a function of semi-major axis for SG (blue) and TQM (red) models which are both scaled to a $10^8 M_{\odot}$ SMBH. In the top panel, 1-d radial height profiles are reflected both across the disk mid-plane (black horizontal dashed line) and to the far side of the schematic SMBH to illustrate the model cross-sections. The orbit of the schematic off-plane orbiter are defined by the semi-major axis ($a$) and inclination ($i$). We assume orbital eccentricity ($e$) $\sim 0$ for inclined orbiters (see text).
\label{fig:diskprofiles}}
\end{figure}\ 

Here we use two plausible AGN disk models \citet[][]{SG}, hereafter SG, and \citet[][]{TQM}, hereafter TQM.  Figure~\ref{fig:diskprofiles} (top panel) compares the 1-d radial disk height profiles of the SG (blue) and TQM (red) models. A schematic SMBH and orbiting star are added to illustrate a disk-crossing orbiter. Fig.~\ref{fig:diskprofiles} (lower panel) compares the 1-d radial density profiles for both models. SG is constructed to match observed AGN optical/UV SEDs and is therefore likely more representative of inner AGN disks. Thus, any AGN displaying a 'Big Blue Bump' is likely well-described by an SG model, at least in the region interior to $10^{4-5}~R_g$, and this is the majority of bright AGN (i.e. QSOs and Seyferts \cite{Netzer15}). TQM is constructed to match inferred mass inflow from star formation and so is likely more representative of outer AGN disks. Nevertheless both models represent plausible AGN disks which typically have larger aspect ratios than a Shakura-Sunyaev disk model \citep{SS73}. From Fig.~\ref{fig:diskprofiles}, NSC stars or stellar remnants with $a <10^{7}~R_{g}$ interact with the disks. At small semi-major axes, ($a \leq 10^{4}~R_{g}$), the Keplerian orbital period ($T_{\rm orb}=2 \pi a^{3/2}G^{-1/2} M_{\rm SMBH}^{-1/2}$) is shortest and $\rho$ is largest, so we anticipate that drag effects are most important in these models at small disk semi-major axes. At larger semi-major axes ($a>10^{4}~R_{g}$), gas density falls quickly and orbital time is long, so we anticipate a less efficient, slower capture process at large semi-major axes in these models. 

\subsection{Drag on disk-crossing orbiters} \label{sec:drag}

When an NSC orbiter plunges through the gas disk (e.g. in the  schematic in Fig.~\ref{fig:diskprofiles}), it experiences a drag force from the gas. For a sufficiently long-lived disk, or a sufficiently small semi-major axis, repeated disk-orbiter interactions tend to reduce both the inclination angle ($i$) and semi-major axis ($a$) of the orbiter so that $\frac {da}{dt},\frac {di}{dt}<0$ and given enough time the orbiter is captured by the disk.  We define disk capture when $(a,i)$ are completely contained within the disk radial height profile. We assume orbiters are subject to zero drag for the portion of their orbit that is outside the disk. For the purposes of the present work we ignore objects on retrograde orbits with respect to the disk gas and we leave discussion of retrograde orbiters to future work. We also ignore the effect of vertical Lindblad resonances \cite{Lubow81,ArtyLubow94} which are likely to accelerate the processes described here. The drag forces experienced by stars and stellar remnants are different and we describe them each below. 

\subsubsection{Geometric Drag} \label{sec:GEOdrag}

When the orbiter surface is well defined, i.e. for every orbiter we consider except black holes, geometric drag provides the primary drag force. We define the geometric drag force ($F_{\rm GEO}$) as \citep{Passy12}:
\begin{equation} \label{eqn:Fdrag-GEO}
 F_{\rm GEO} = \frac{1}{2}C_{\rm d} \, (4\pi r_*^2) \, \rho_{\rm disk} {\rm v_{rel}}^2   
\end{equation}
where $r_*$ is the orbiter surface radius, $\rho_{\rm disk}$ is the local disk density, and ${\rm v_{rel}}$ is the relative velocity between the orbiter and the disk.  We assume the drag coefficient, $C_d=1$, appropriate for a spherical star or stellar remnant (in general it is a factor of order unity at high Reynolds number, appropriate to AGN disks). The action of the drag  causes loss of kinetic energy and momentum. 
\footnote{Kinetic energy losses from the orbit are converted to thermal energy in the disk, adding to local turbulence and viscosity. This will be a topic of future work.}

\subsubsection{Bondi-Hoyle-Lyttleton Drag} \label{sec:BHLdrag}

Black holes do not have a solid surface so geometric drag is not appropriate for this case. However, gas will flow around the sBH to create a Bondi-Hoyle-Lyttleton (BHL) drag shock tail. This tail acts to gravitationally slow the passage of the sBH. The drag force due to BHL accretion ($F_{\rm BHL}$) is \citep[e.g.][]{Ostriker99,Antoni19}:
\begin{equation} \label{eqn:Fdrag-BHL}
    F_{\rm BHL} = \dot M_{\rm BHL} \, \rm{v}_{\rm rel}
    = \frac{4 \pi {\it G}^{2} {\it M}_{\rm BH}^{2} \rho_{\rm disk}}{\rm{v}^{2}_{\rm rel}}
\end{equation}
where $\dot{M}_{\rm BHL}$ is the usual expression for the Bondi mass accretion rate and $M_{\rm BH}$ is the mass of the sBH. For computational efficiency we only consider a single $M_{\rm BH}=10M_{\odot}$. Because of the $M_{\rm BH}$ dependence we expect that AGNs will be more efficient at capturing more massive back holes such as the ones that LIGO-Virgo has detected. $F_{\rm BHL}$ has a $1/\rm{v}_{\rm rel}^2$ dependence instead of the $\rm{v}_{\rm rel}^2$ in $F_{\rm GEO}$. As a result, sBHs experience greater drag at small inclination angle ($i$)  (where $\rm{v}_{\rm rel}$ is always small) and at a large semi-major axis $a$ (where Keplerian velocities are small). We will ignore the effects of BHL accretion ($\dot{M}_{\rm BHL}$) on the sBH mass during passage through the disk. However, the EM luminosity associated with disk crossing sBH, $L_{\rm BHL}=\eta \dot{M}_{\rm BHL} c^{2}$, is likely to be most luminous at small $i$, where $\dot{M}_{\rm BHL}$ is largest and assuming $\eta$, the radiative efficiency of the accreting gas, is a few percent (see also Section \ref{sec:EM}).

\subsection{Capture-time for inclined orbits} \label{sec:capturetime}

Consider a single orbiter of mass $M$ and radius $r_*$ on a prograde Keplerian orbit of semi-major axis, $a$, and inclination angle, $i$ with respect to the disk plane. In the simplest case, if we assume that each passage results in a negligible change in $a$ (i.e. $\frac {da}{dt} \approx 0$ per orbit), then $\frac {di}{dt}<0$ due to drag until the inclination angle $i \leq i_{\rm crit}$ where 
\begin{equation}\label{eqn:icrit}
    i_{\rm crit} = \arcsin \left(\frac{H}{2a \sin i}\right)
\end{equation}
where $H$ is the disk height and the orbiter is then embedded in the disk. For computational simplicity, and to derive an analytic expression for the capture time which may be valid in limited circumstances, we initially assume that $\frac {da}{dt}=0$. Our analytic estimate of the time taken for the disk to capture the orbiter ($T_{\rm cap}$) is frequently an upper limit, i.e. $T_{\rm cap}<T_{\rm orb} \frac{dv_{\rm z}}{v_{\rm z}}$ where $dv_{\rm z}$ is the change in the orbital velocity component perpendicular to the disk plane ($v_{\rm z}$) and $T_{\rm orb}$ is the Keplerian orbital period. For orbiters undergoing geometric drag therefore, 
\begin{equation} \label{eqn:tcap-GEOprelim}
    T_{\rm cap~{GEO}} < \frac{4}{3\pi} \left( \frac{\rho_*}{\rho_{\rm disk}}\right) \left(\frac{r_*}{a}\right) \left(\frac{1}{\sin i} \right)\frac{1}{\arcsin \left(\frac{H}{2a \sin i}\right)} T_{\rm orb}
\end{equation}
for a star of density $\rho_*$ and semi-major axis $r_*$ and where the arcsin term corresponds to the fraction of the orbit spent in the disk. The equivalent upper limit for disk capture of sBH is 
\begin{equation} \label{eqn:tcap-BHLprelim}
    T_{\rm cap~{BHL}}< \frac{{\rm{v_{rel}}}^4 \sin^3i}{8\pi^2 a G^2M_{\rm BH}  \, \rho_{\rm disk} \arcsin \left(\frac{H}{2a \sin i}\right)} T_{\rm orb}.
\end{equation}
 
However, for small $a$ and high $\rho_{\rm disk}$, we should \emph{expect} $\frac {da}{dt}$ to be signficantly non-zero. In this case, we should allow $\frac {da}{dt}<0$ and find the change, due to drag, in orbital velocity components perpendicular ($dv_{\rm z}$) and parallel ($dv_{\rm \theta}$) to the disk plane for each passage. The orbital inclination after the ${\rm j}$th passage through the disk is then
\begin{equation} \label{eqn:ij}
    i_{\rm{j+1}} = \rm \arctan \left(\frac{v_{\rm z_j}+dv_{\rm z_j}}{v_{\rm \theta_j}-dv_{\rm \theta_j}}\right).
\end{equation}
The loss of kinetic energy due to drag induced work, 
 \begin{equation} \label{eqn:work}
     W_{\rm drag,j} = F_{\rm drag,j} R_{\rm arc,disk,j} = F_{\rm drag,j} \times {2a_{ \rm j}} \arcsin \left(\frac{H}{2a_{ \rm j} \sin i_{ \rm j}}\right)
 \end{equation}
where $R_{\rm arc,disk,j}$ is the arc of the orbit embedded in the disk on a single pass, goes directly into potential energy, as the orbiter moves deeper into the the SMBH potential well.
In Eqn.\ref{eqn:work} we can substitute $F_{\rm drag}$ with $F_{\rm BHL}$ for sBH and $F_{\rm GEO}$ for all other orbiters, and ($a_{\rm j},i_{\rm j}$) are the orbital semi-major axis and inclination angle on the ${\rm j}$th pass.  So the change in semi-major axis ($\Delta a_{\rm j}$) after the ${\rm j}$th disk passage is
\begin{equation} \label{eqn:dela}
    \Delta a_{\rm j} = \frac{GM_{\rm SMBH}}{2W_{\rm drag, j}}
\end{equation}
Writing the orbital semi-major axis after the ${\rm (j+1)}$th passage through the disk as 
\begin{equation} \label{eqn:aj}
    a_{\rm j+1}=a_{\rm j}-\Delta a_{\rm j}
\end{equation}
the final time to capture is 
\begin{equation} \label{eqn:tcap}
    T_{\rm cap} = \sum_{\rm j=0}^{2n(i_{0}, a_{0})} \frac{T_{\rm orb}(a_{\rm j})}{2} = \sum_{\rm j=0}^{2n} \frac{\pi a_{\rm j+1}^{3/2}}{\sqrt{G M_{\rm SMBH}}}
\end{equation}
where $n=\frac{\rm j_{crit}}{2}$ is the number of orbits to achieve capture ($i_{\rm j}\leq i_{\rm crit}$). In general, $n$ will depend on $i_{0}$, the initial orbital inclination angle, and $a_{0}$, the initial semi-major axis, and we must find $n$ and $T_{\rm cap}$ by numerical integration over a specific disk model.

\section{Results} \label{sec:results}

Here we discuss the time taken for our two AGN disk models to capture various stellar types and sBH as a function of $i_{0}$ and $a_{0}$. 

\subsection{Geometric Drag}\label{sect:Geometric_stars}

\begin{figure}
\includegraphics[width=\linewidth]{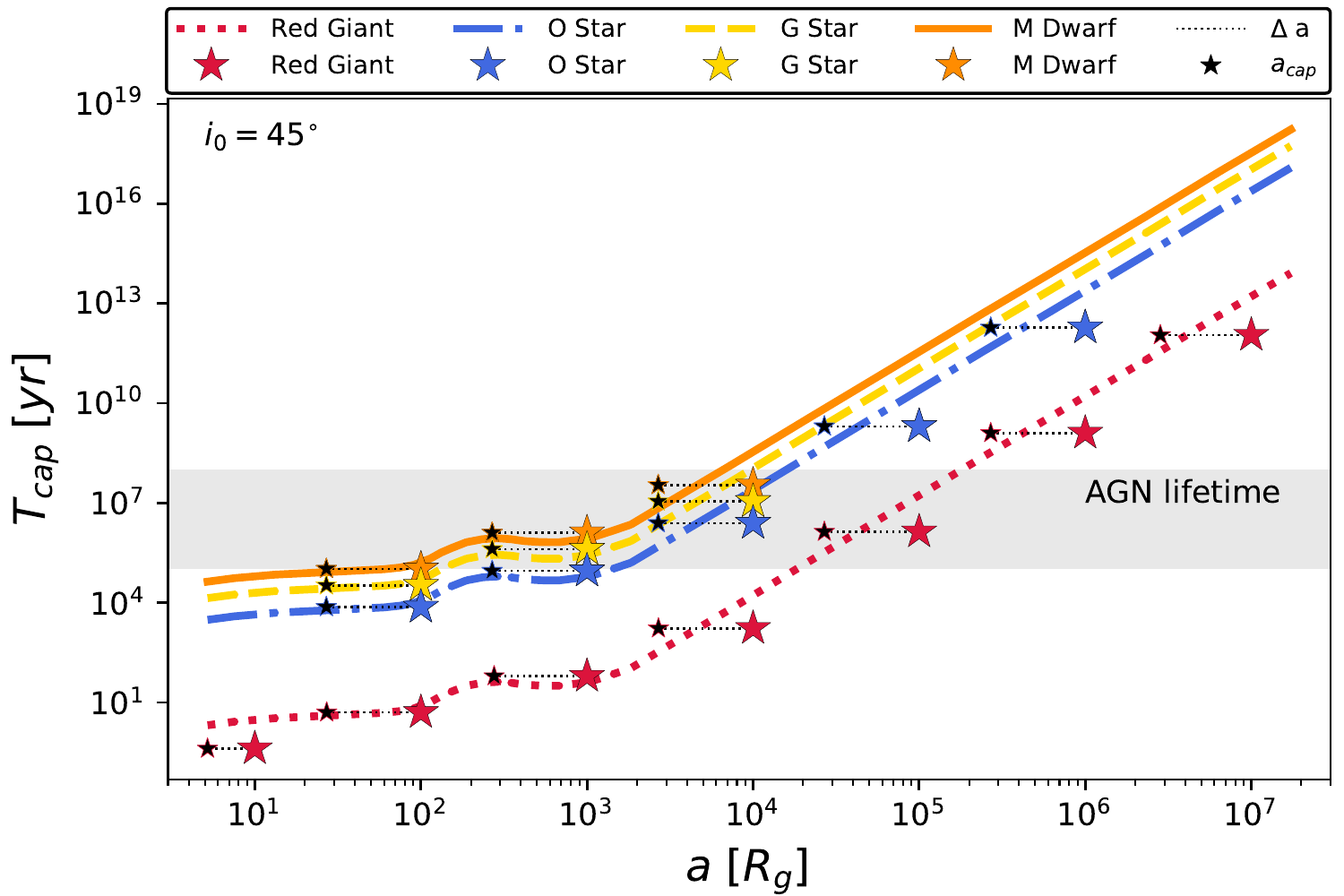}
\includegraphics[width=\linewidth]{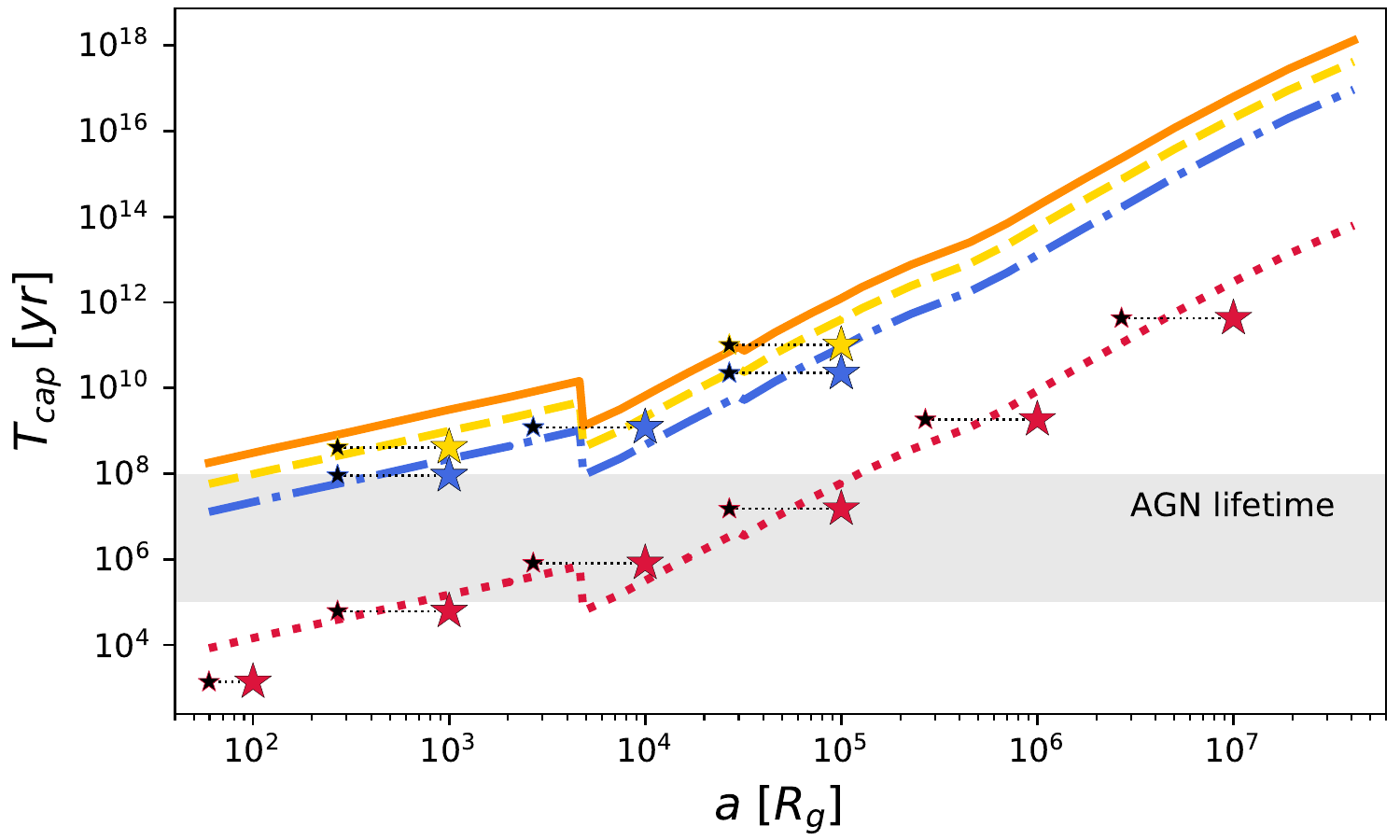}
\caption{Capture-Time ($T_{\rm cap}$) due to geometric drag as a function of initial orbital semi-major axis ($a_{0}$) for the four stellar objects from Table~\ref{table:properties} for a SG disk model (top panel) and TQM disk model (bottom panel). Each star has a Keplerian prograde orbit with $i_{0}=45^{\circ}$. Curves correspond to the analytic estimate of $T_{\rm cap}$ from eqn.~\ref{eqn:tcap-GEOprelim}. Large coloured star and small black star symbols indicate the initial $a_{0}$ and final $a$ for each stellar type after $T_{\rm cap}$ (which is read off the vertical axis of the plot), found from numerical integration of eqn.~\ref{eqn:tcap}. Grey band corresponds to a plausible range of AGN disk lifetimes (see text). The largest stellar radii are captured most quickly for a given model, but the capture times for the SG model are always shorter, due to higher gas densities; thus, an SG disk captures all stars with $a<10^{4}~R_{g}$ at $i_{0}=45^{\circ}$, while a TQM disk captures only Red Giants at most radii, and captures other stars only at very small disk radii.
\label{fig:stellar-Tcap_a}}
\end{figure}

\begin{figure}
\includegraphics[width=\linewidth]{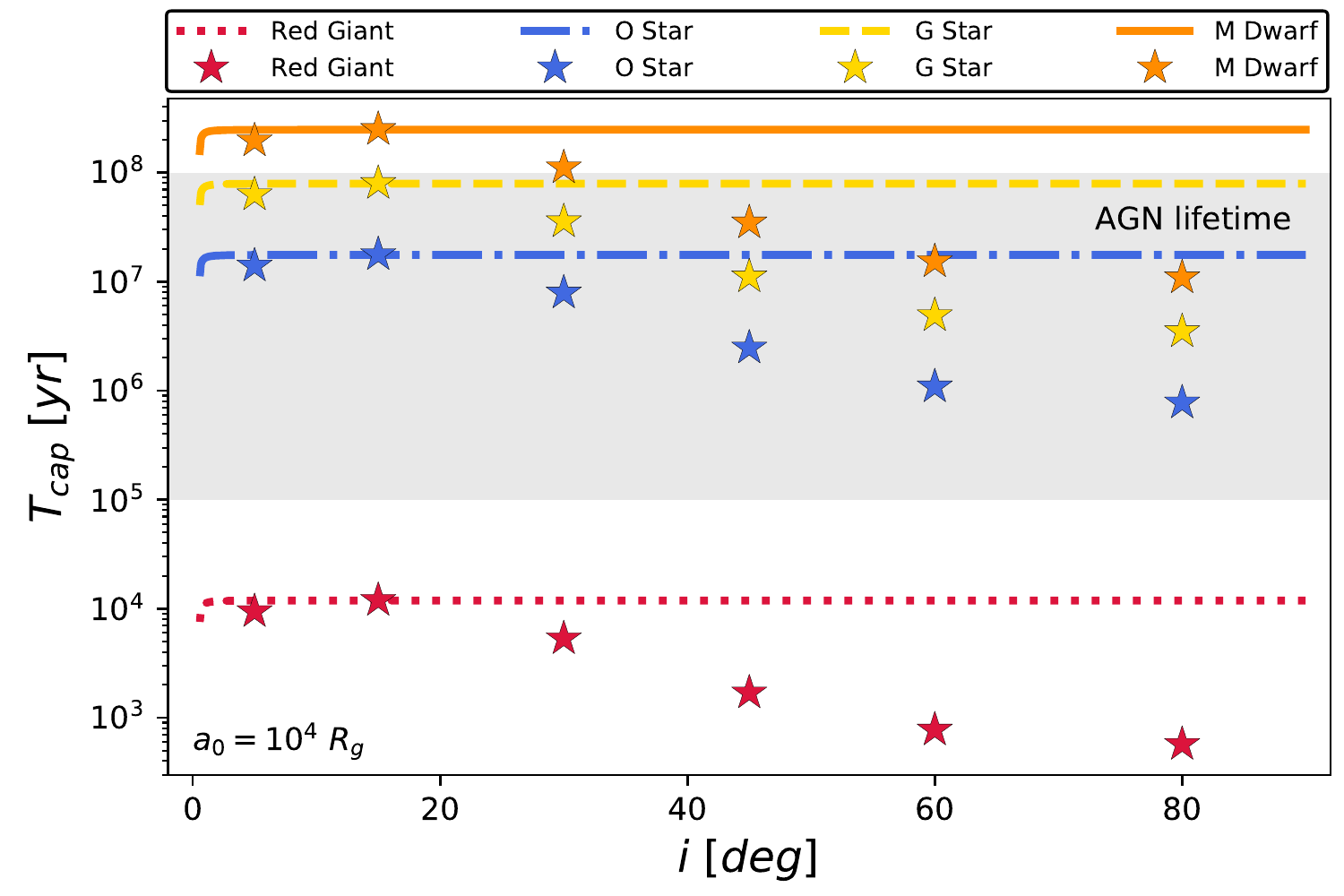}
\includegraphics[width=\linewidth]{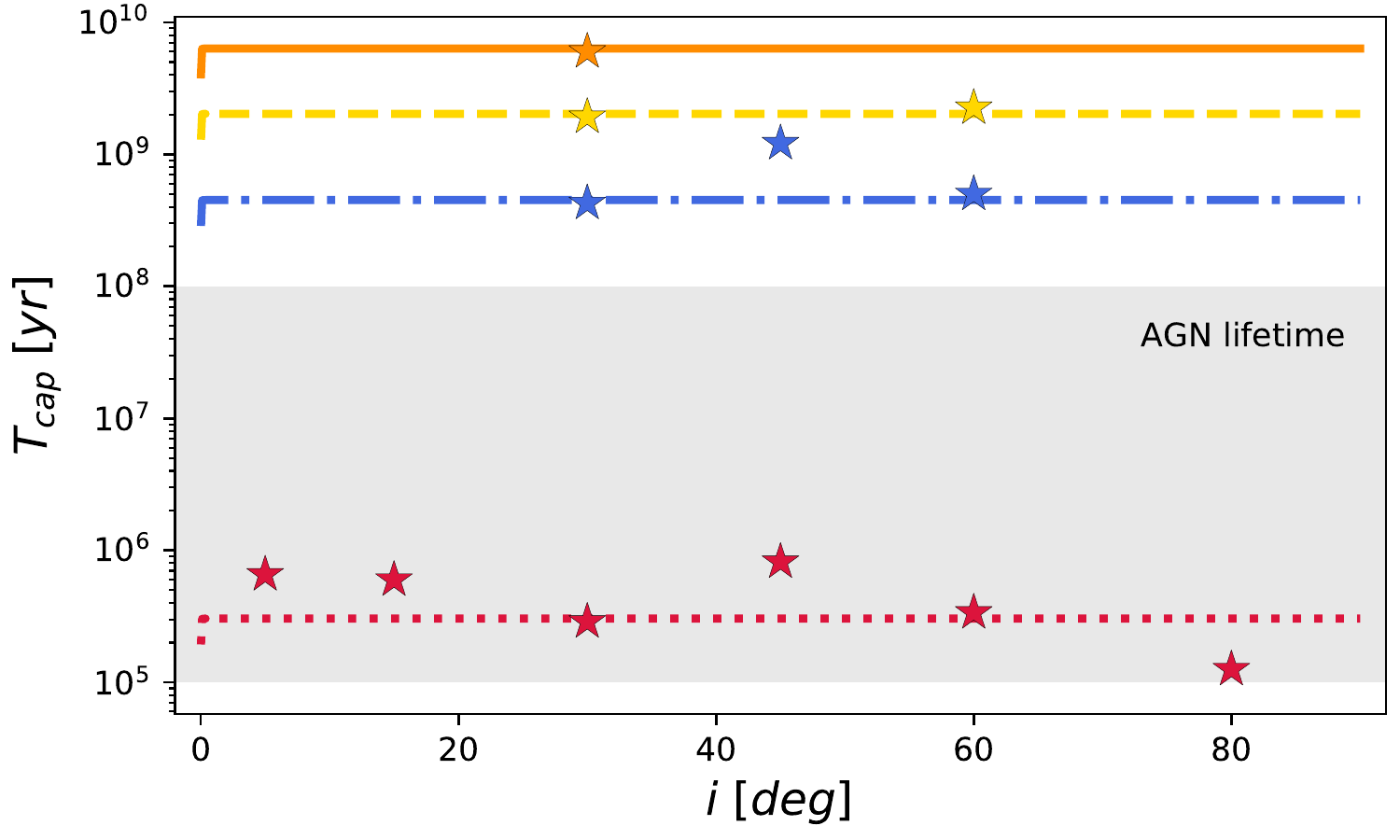}
\caption{Capture time $T_{\rm cap}$ due to geometric drag as a function of initial orbital inclination ($i_{0}$) with notation as for Fig.~\ref{fig:stellar-Tcap_a}. We assumed Keplerian prograde orbits with $a_{0}=10^{4}~R_{g}$. Curves correspond to the analytic estimate for $T_{\rm cap}$ from eqn.~\ref{eqn:tcap-GEOprelim} and star symbols indicate $T_{\rm cap}$ from numerical integration of eqn.~\ref{eqn:tcap}. Conservation of energy causes
higher orbital inclinations to produce large changes in $a$, as
the number of orbits needed to reach $i_{crit}$ increases. This results in a lower $T_{\rm cap}$ for higher inclinations
at the same initial semi-major axis. Changing $a$ also causes
the anomalies seen in the TQM panel, due to orbiters crossing the (unphysical) density discontinuity at approximately $5 \times 10^{3}~R_{g}$.
\label{fig:stellar-Tcap_i}} 
\end{figure}

Fig.~\ref{fig:stellar-Tcap_a} shows $T_{\rm cap}$ as a function of  $a_{0}$ for stars on Keplerian prograde orbits assuming  $i_{0}=45^{\circ}$ interacting with a SG disk (top panel) and a TQM disk (bottom panel) respectively. The stars have properties as in Table~\ref{table:properties}. Coloured curves correspond to the analytic approximation for $T_{\rm cap}$ from eqn.~\ref{eqn:tcap-GEOprelim} for each stellar type. Large colored star symbols and small black star symbols indicate $a_{0}$ and final $a$ for a given $T_{\rm cap}$ (read off vertical axis), from numerical integration of eqn.~\ref{eqn:tcap}. The grey horizontal band corresponds to a fiducial range of AGN disk lifetimes (0.1-100~Myr), which is consistent with values inferred from observations \citep{Haehnelt93,King15,Schawinski15}. For most choices of orbital initial conditions, $T_{\rm cap}$ from eqn.~\ref{eqn:tcap} is less than or equal to the value from the analytic approximation of eqn.~\ref{eqn:tcap-GEOprelim}, so our analytic form is a convenient upper limit. 

Fig.~\ref{fig:stellar-Tcap_i} shows $T_{\rm cap}$ as function of $i_{0}$ for different stellar types assuming $a_{0}=10^{4}~R_{g}$. The curves in Fig.~\ref{fig:stellar-Tcap_i} correspond to the analytic expression in eqn.~\ref{eqn:tcap-GEOprelim}, which is very weakly dependent on $i$, hence the flatness of the curves. However, Fig.~\ref{fig:stellar-Tcap_i} also shows that numerical integration is required for highly inclined orbiters, and for disk models with highly variable radial height profiles (e.g. TQM). Counter-intuitively, highly inclined stellar orbiters are captured faster by the SG disk since relative velocity is largest for high $i_{0}$ and therefore $F_{\rm GEO}$ is largest. The TQM disk model is less efficient than the SG disk model at star capture at $a_{0}=10^{4}~R_{g}$ for all stellar types, with only Red Giants ending up within the disk for plausible AGN lifteimes. For simplicity we assume no change in stellar mass due to disk crossings. For red giants in particular this is likely incorrect since \citet{Kieffer16} find that RG can lose up to 10\% of their mass during a disk crossing (though for column densities at least two orders of magnitude greater than any considered here). Further complicating the expected capture time of giants, \citet{Armitage96} find that mass loss due to repeated impacts with the disk leads to an expansion of the star and a subsequent increase in cross section. Stars with a larger surface area are captured by the disk faster since they experience a stronger drag from eqn. \ref{eqn:Fdrag-GEO}, and therefore including the mass loss in our calculations could \emph{reduce} capture time for giants in both SG and TQM. In addition Armitage finds that a main sequence star can survive repeated disk passages while RG are destroyed much more quickly; this could explain the observed absence of RG in the nuclear star cluster of the Milky Way.
Some variations in $T_{\rm cap}$ as a function of $i$ for the TQM disk appear to be due to changes in $a$ across the sharp (unphysical) density and scale height changes at $a \sim 5 \times 10^{3}~R_{g}$.

From the top panels of Fig.~\ref{fig:stellar-Tcap_a} and Fig.~\ref{fig:stellar-Tcap_i}, stars on prograde orbits with $a_{0}\leq 10^{3}~R_{g}$ will all end up within SG-type AGN disks on short timescales. For AGN lifetimes of $\sim 1$Myr, most stars within $a<10^{4}~R_{g}$ will end up within SG-type AGN disks. Assuming a $\sim 10^{6}M_{\odot}/\rm{pc}^{3}$ uniform stellar density nuclear star cluster at $a<0.1$pc from the SMBH \citep{Antonini14}, we should expect $O(10^{3})$ stars with $a<10^{4}~R_{g}$. So, several hundred stars in the core of an NSC will be captured by a SG-type AGN disk over a fiducial $\sim 1$Myr lifetime. Conversely, if most AGN disks are more like TQM than SG, we should expect few NSC stars to be captured by AGN disks. Hence, the rate of supernovae, tidal disruptions and other stellar EM signatures associated with embedded stellar objects in AGN disks is a direct probe of mean AGN disk properties (see also below). In both disk models, stars are most quickly captured by the disk at small semi-major axes where $T_{\rm orb}$ is small.

\subsection{Bondi-Hoyle-Lyttleton Drag}\label{Bondi-Hoyle-Lyttleton Drag}

\begin{figure}
\includegraphics[width=\linewidth]{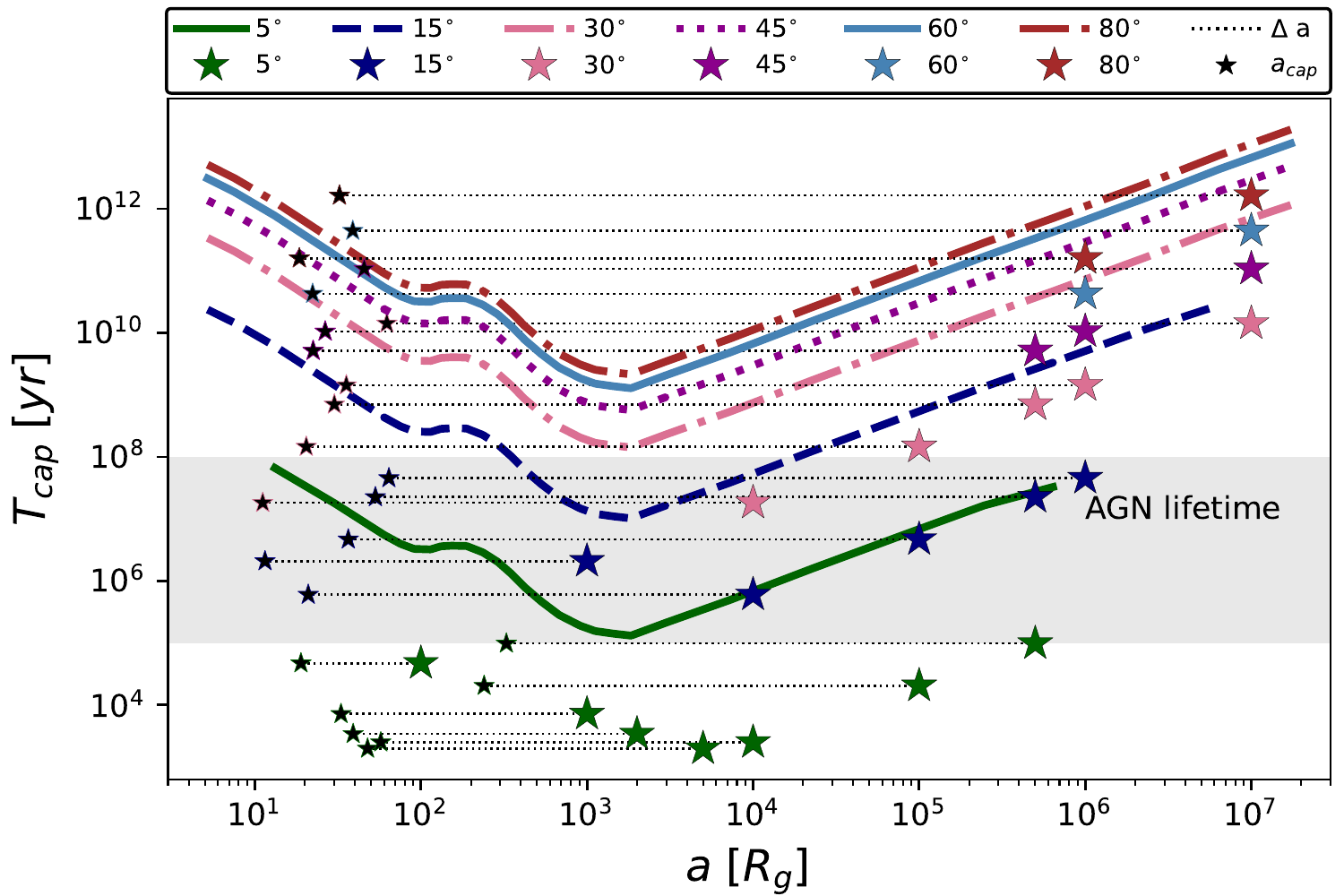}
\includegraphics[width=\linewidth]{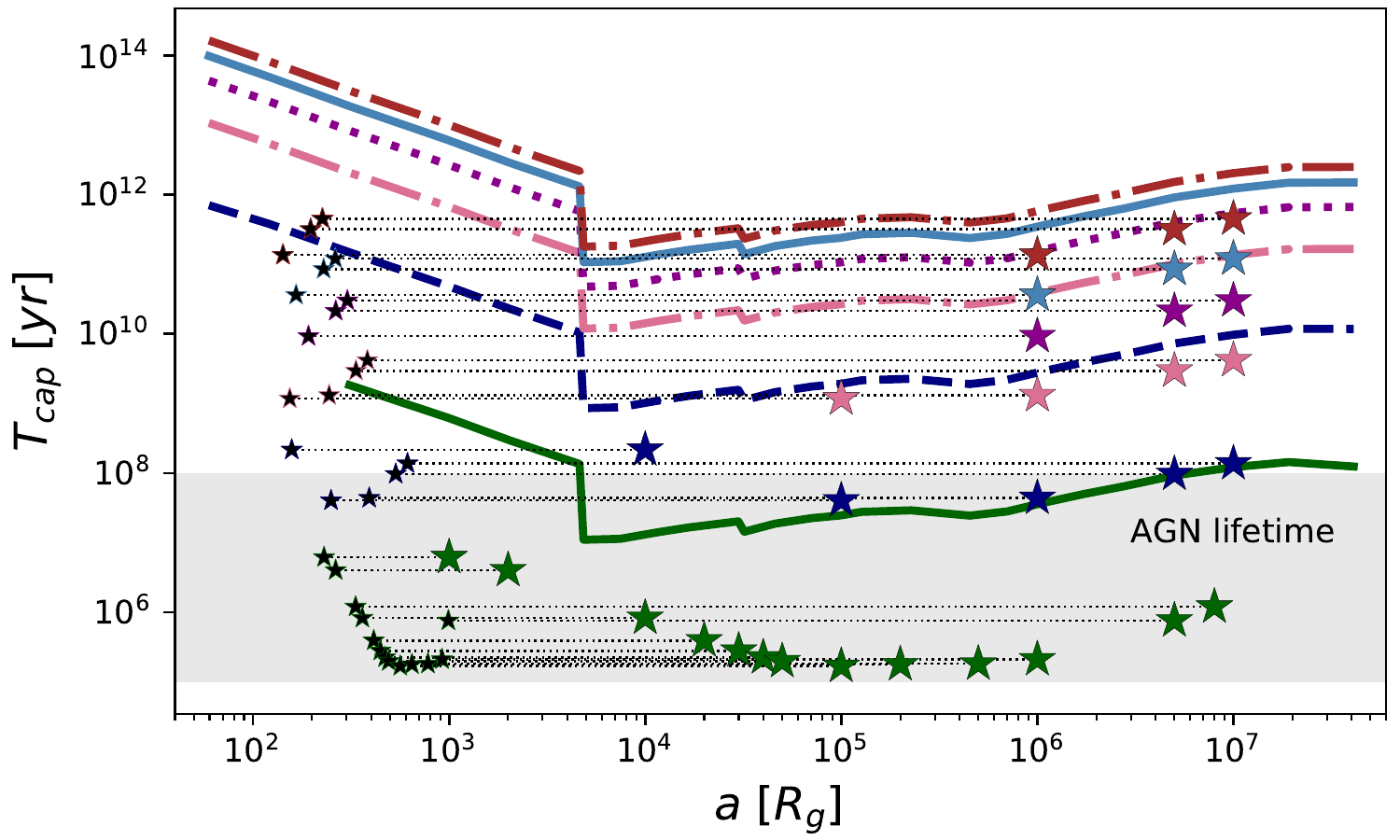}
\caption{As Fig.~\ref{fig:stellar-Tcap_a} but for ($T_{\rm cap}$) due to BHL drag as a function of $a_{0}$ for a sBH (10$M_{\odot}$) on an inclined orbit w.r.t. a SG disk model (top panel) and TQM disk model (bottom panel). The sBH are on Keplerian prograde orbits with $i_{0}=5^{\circ},15^{\circ},30^{\circ},45^{\circ},60^{\circ},80^{\circ}$ respectively (each with an associated colour). Curves correspond to analytic estimates for $T_{\rm cap}$ from eqn.~\ref{eqn:tcap-BHLprelim}. As in Fig.~\ref{fig:stellar-Tcap_a}, large coloured star symbols and small star symbols correspond to $a_{0}$ and final $a$ at a given $T_{\rm cap}$ (which is read off the vertical axis), found from numerical integration of eqn.~\ref{eqn:tcap}. Because sBH on high inclination orbits have initially very high relative velocities, their orbits are barely perturbed by BHL drag, and they spend most of their orbital decay at $\sim$ constant $a$ and slowly decreasing $i$. This continues until the combination of $(a,i)$ produces a ${\rm v_{rel}}$ that causes large perturbations per disk-crossing, leading to rapid capture. Thus eqn.~\ref{eqn:tcap-BHLprelim} is a reasonable guide for $T_{\rm cap}$ at high $i$, but becomes an increasingly poor approximation at low $i$, where the time to reach the critical plunge conditions is small. Note also that the sBH are always captured at small radii, where the aspect ratio ($H/a$) of the respective disk model sharply increases--hence the importance of accounting for the large $\Delta a$.
\label{fig:sBH-Tcap_a}}
\end{figure}

\begin{figure}
\includegraphics[width=\linewidth]{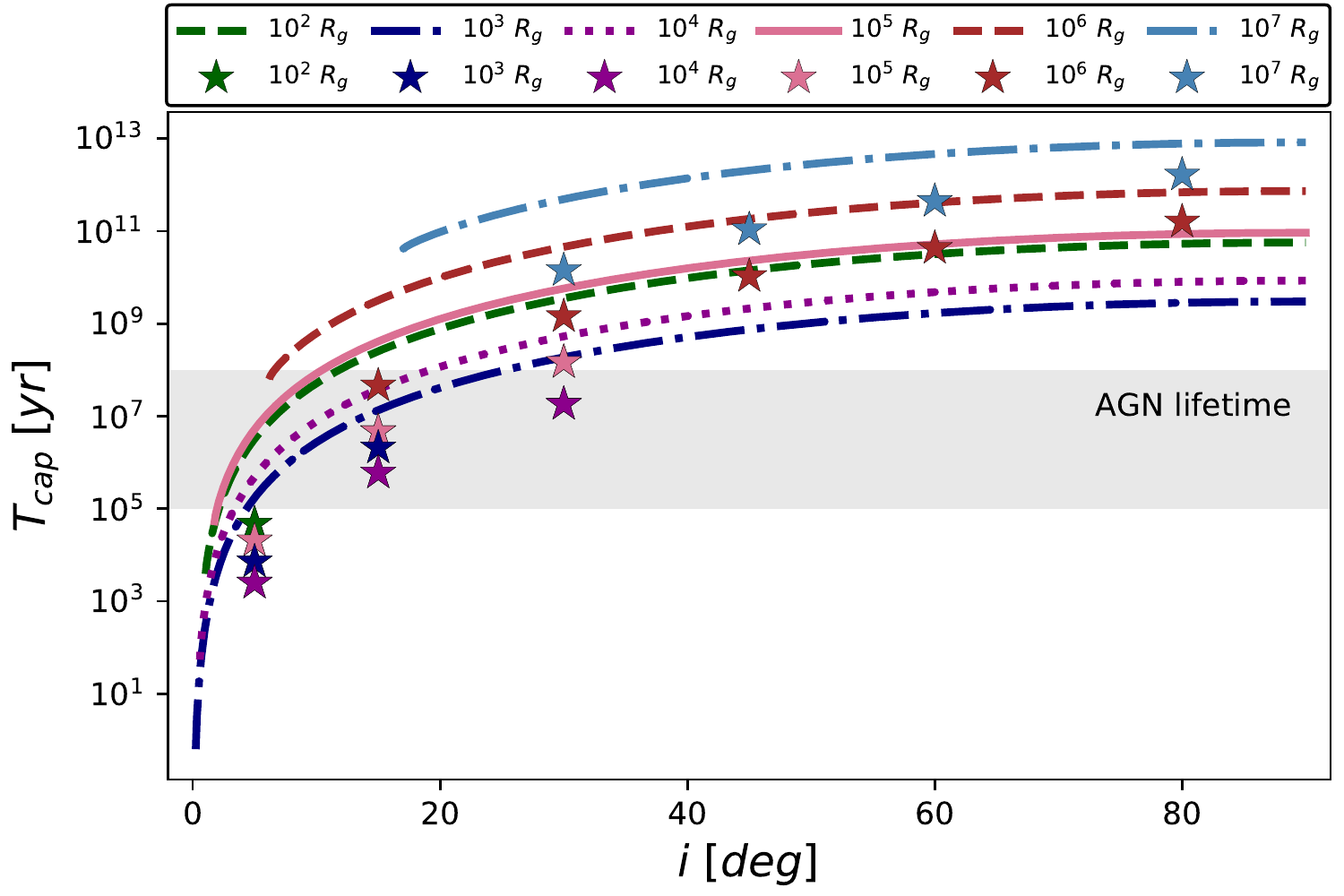}
\includegraphics[width=\linewidth]{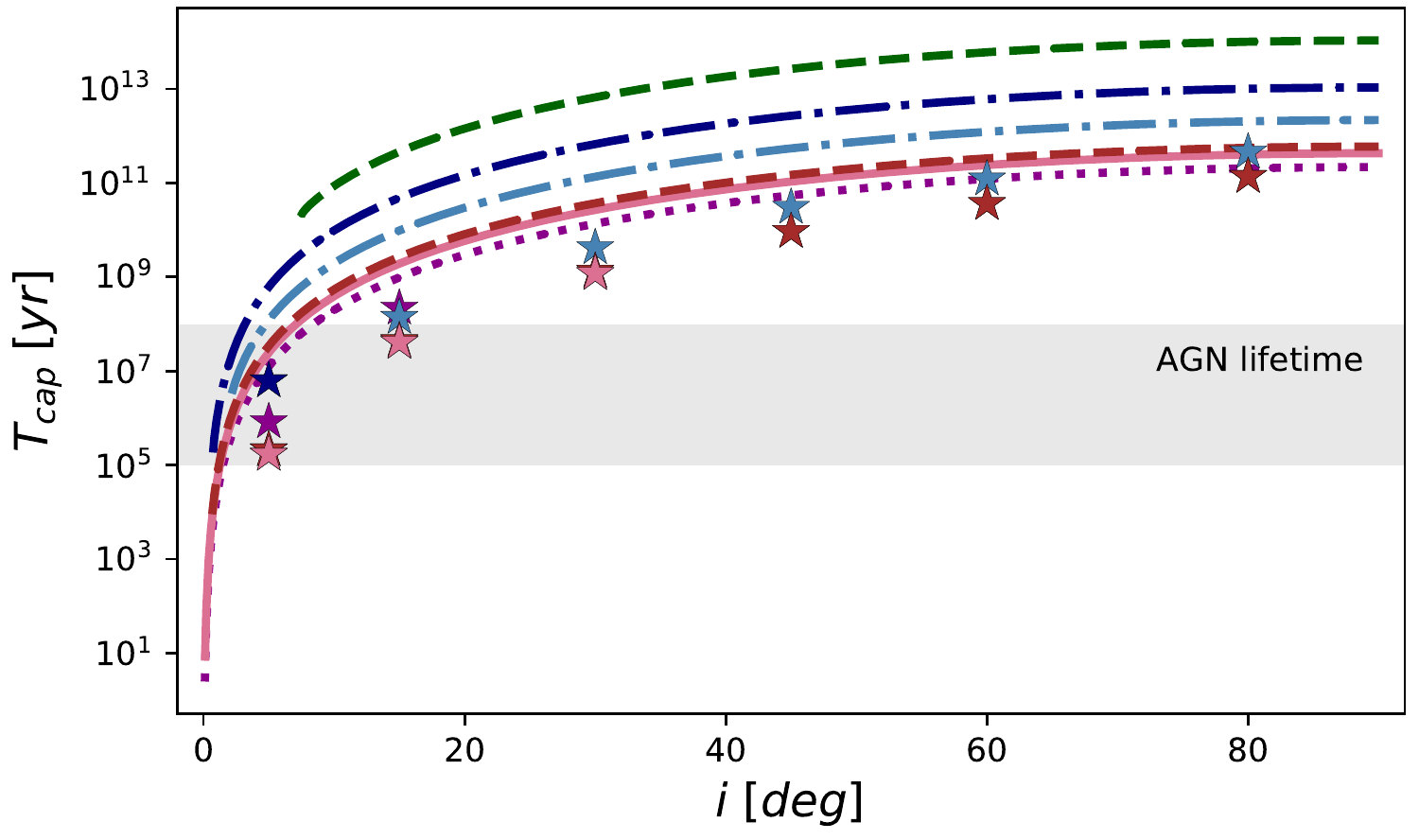}
\caption{As Fig.~\ref{fig:stellar-Tcap_i} except $T_{\rm cap}$ due to BHL drag as a function of $i_{0}$ for sBH for a range of $a_{0}$. Curves correspond to analytic estimates of $T_{\rm cap}$ at each $a_{0}$ from eqn.~\ref{eqn:tcap-BHLprelim} and the curves end at $i_{\rm crit}$ for each $a_{0}$. Star symbols correspond to $T_{\rm cap}$ found by numerical integration of eqn.~\ref{eqn:tcap}. For $i_0<15^{\circ}$, sBH may be captured by either SG or TQM disks, though SG captures them from a wider range of $a_0$; only long-lived SG disks can capture sBH from $i_0\gtrsim 15^{\circ}$.
\label{fig:sBH-Tcap_i}}
\end{figure}

Figure~\ref{fig:sBH-Tcap_a} shows $T_{\rm cap}$ as a function of semi-major axis ($a$) for sBH on Keplerian prograde orbits with initial inclination angles spanning $i_{0}=[5^{\circ},80^{\circ}]$ interacting with a SG disk (top panel) and a TQM disk (bottom panel). The sBH have properties as in Table~\ref{table:properties}. Coloured curves correspond to the analytic expression in eqn.~\ref{eqn:tcap-BHLprelim} for each choice of $i_{0}$. For $i_{0}=5^{\circ},15^{\circ}$, $T_{\rm cap}$ analytic curves are cut off at very small and very large semi-major axes because the value of $i_{crit}$ rises in those regions with the aspect ratio of the disk. Those curves are thus already fully embedded in the disk. Large star symbols indicate starting semi-major axis ($a_{0}$) for sBH with colour indicating $i_{0}$, and are connected by horizontal lines to small black star symbols, indicating the sBH semi-major axis at $T_{\rm cap}$ from numerical integration of eqn.~\ref{eqn:tcap}. 

From Fig.~\ref{fig:sBH-Tcap_a}, several things stand out. First, sBH with inclination angles $\leq 15^{\circ}$ will be captured by an SG-type AGN disk for most of the disk ($10^{2}~R_{g}<a<10^{6}~R_{g}$). For a uniform sBH distribution, this corresponds to a moderate fraction of the entire NSC sBH population. Second, there is a quite remarkable plunge in semi-major axis to small values ($\leq 2 \times 10^{2}~R_{g}$ for SG; $\leq 10^{3}~R_{g}$ for TQM) during orbital capture. This plunge suggests that the BHL drag runs away, as $a$ decreases and brings the sBH into denser regions in the inner disk. One consequence is that O($10\%$) of sBH in an NSC could be delivered in large numbers to the innermost AGN disk over a plausible AGN disk lifetime, with significant implications for the LIGO-Virgo merger rate as well as the build up of massive (IMBH) merger products in the inner disk (see \S\ref{sec:discussion} below). Third, by contrast with stars, disk capture can be surprisingly efficient for sBH at large disk semi-major axes and lower inclinations due to the relative velocity dependence of $F_{\rm BHL}$. Fourth, because sBH on high inclination orbits have initially very high relative velocities, their orbits are barely perturbed by BHL drag, and they spend most of their orbital decay at $\sim$ constant $a$ and slowly decreasing $i$. This continues until the combination of $(a,i)$ produces a ${\rm v_{rel}}$ that causes large perturbations per disk-crossin, leading to rapid capture. Thus eqn.~\ref{eqn:tcap-BHLprelim} is a reasonable guide for $T_{\rm cap}$ at high $i$, but becomes an increasingly poor approximation at low $i$, where the time to reach the critical plunge conditions is small. Fifth, merging black holes with modest kicks directed significantly out of the disk will likely lead to the removal of the merger product from the disk, and we consider disk recapture for fiducial AGN disk lifetimes in \ref{sec:denseAGN}.

Fig.~\ref{fig:sBH-Tcap_i} shows $T_{\rm cap}$ as a function of $i_{0}$  for sBH with $a_{0}=[10^{2},10^{7}]~R_{g}$ for the SG disk model (top panel) and TQM (bottom panel). Curves correspond to the analytic expression for $T_{\rm cap}$ from eqn.\ref{eqn:tcap-BHLprelim} and these have smallest values at the densest semi-major axes in the two models. From the top panel it is clear that sBH across a wide range of semi-major axes ($\sim 10^{2}-10^{6}~R_{g}$) can be captured efficiently by SG-type AGN disks for $i_{0} \leq 15^{\circ}$. From Fig. \ref{fig:sBH-Tcap_a}, we expect sBH to be captured for $i_{0}\leq30^{\circ}$ for semi-major axes near $10^{4}~R_{g}$. sBH orbits with $i_{0}\leq15^{\circ}$ can be captured by a TQM-type disk within plausible AGN disk lifetimes.

\begin{figure}
\includegraphics[width=\linewidth]{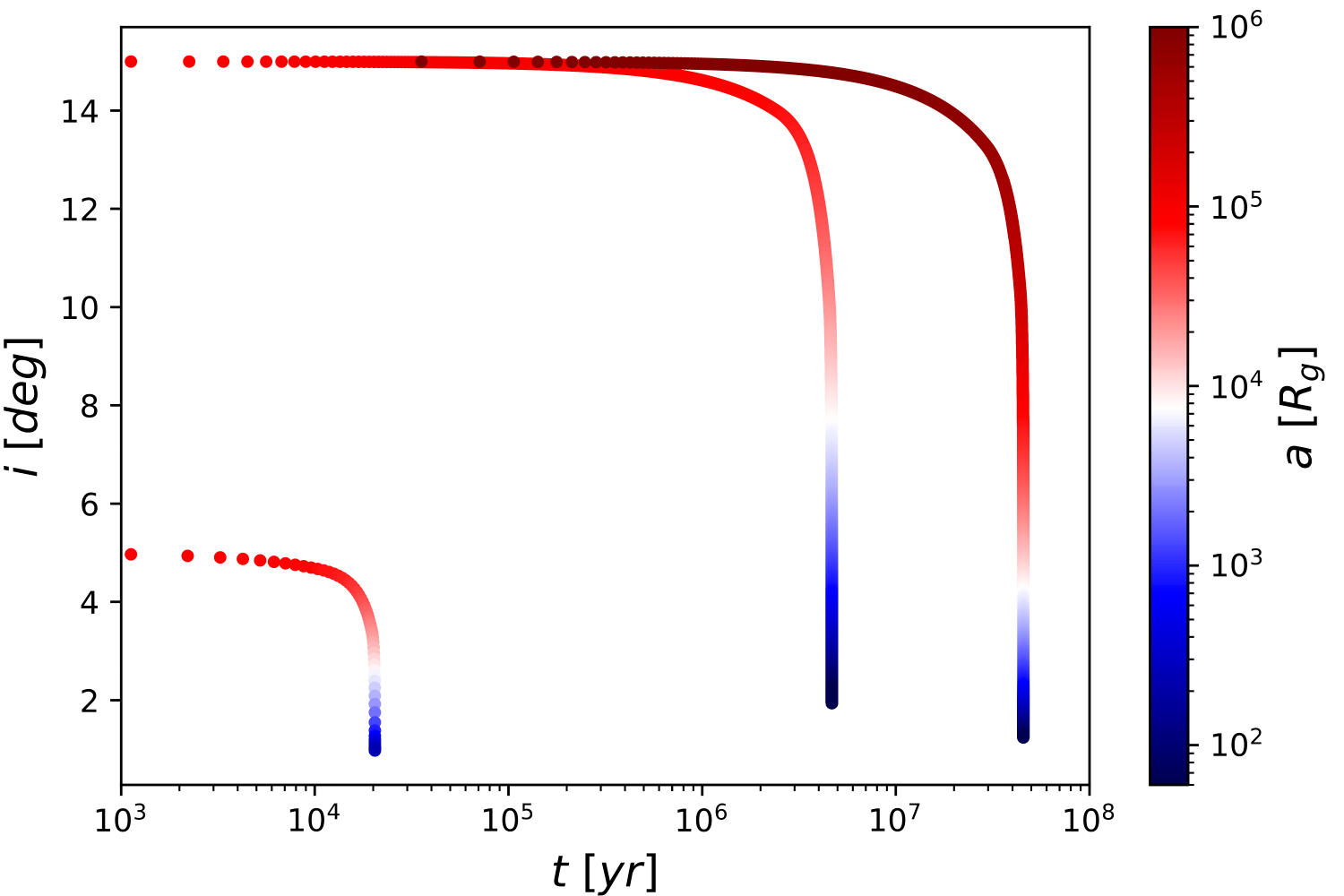}
\caption{Change in orbital inclination angle as a function of time for sBH with $i_{0} = 5^{\circ}$ and $15^{\circ}$, and $a_{0} = 10^5$ and $10^6 R_{g}$, respectively, interacting with an SG disk model (compare with Fig.~\ref{fig:sBH-Tcap_a} top panel). Note the behavior of $\sim$constant $a$ with slowly decreasing $i$, prior to a critical ${\rm v_{rel}}$ which induces a a rapid plunge. The greatest potential for EM counterparts of disk-crossing lies when $di/dt$ is greatest.
\label{fig:orbit_change}}
\end{figure}

We demonstrate the remarkable collapse in semi-major axis of captured sBH in Fig. \ref{fig:orbit_change}. Here we plot the change in orbital inclination as a function of time for an sBH with $i_{0}=[5^\circ, 15^\circ]$ interacting with an SG disk. The curves are color coded in terms of the change in semi-major axis with $a_{0}=[10^5,10^6]~R_{g}$. For an orbit initially inclined at $5^\circ$ and starting at $10^5 R_{g}$, the semi-major axis is reduced by three orders of magnitude, while for the same $a_{0}$ but for $i_{0}=15^\circ$, $a$ decreases by four orders of magnitude. EM counterparts to this plunge in $a$ may be detectable (see \S\ref{sec:EM}). 

\section{Discussion}
\label{sec:discussion}

Orbital capture by AGN disks and the implications that follow from that (mergers, build up of IMBH) are strongly dependent on the assumed disk gas properties. While we only consider a single fiducial set of parameters for each disk model, we can reasonably infer that higher mass accretion rate AGN will produce shorter capture times, while lower mass accretion rate AGN will produce longer capture times. Our results should scale with $\Sigma=2\rho_{disk}H/a$, the surface mass density of the AGN disk (as long as the mass accretion rate is high enough to form a disk). Due to the non-linear dependence of capture time on disk scale height, we cannot specify a precise relationship of capture time to $\Sigma$ and note this would be a valuable future study. Here we compare our results with those of other groups and we discuss the implications of our results for GW and EM detectors as well as prospects for determining the properties of AGN disks from GW and EM observations.

\subsection{Comparison with recent work}\label{sec:comparison}

Other groups have worked on aspects of the same problem from a range of different angles, including semi-analytic work \citep{MacLeod20}, and high resolution N-body simulations using $\phi$GRAPE \citep{Just12,Kennedy16,Bekdaulet2018}.
In the latter case, the authors adopt two disk models, one with constant height and the other with varying height, which are variants of classic  \citep{SS73, NovikovThorne73} thin-disk models, characterized by a Keplerian rotation curve and with a disk lifetime of 100 Myr. While they adopt a dissipative force equivalent to geometric drag, unlike our single-particle analytic approach these authors simulate the interaction of a collection of Sun-like stars with their model disk \citep{Just12,Kennedy16}. \citet{Kennedy16} show that particle orbital eccentricity quickly decays to $e \sim 0$, which
lends support to our simplifying assumption of circular orbits above. Results from \citep{Kennedy16} also show that prograde circular orbits are better captured in the model with varying disk thickness as a function of semi-major axis, whereas the disk model of constant height has a low number of stars that are captured at initial inclinations higher than $30^\circ$. In \citet{Bekdaulet2018}, the work in \citet{Just12} and \citet{Kennedy16} is extended to include the properties of the nuclear stellar disk and the effects of tidal disruption, and they expect a rate of $\sim 140$ stars captured by a gaseous disk per $\sim 1$Myr of evolution.

In our case, if we assume a uniform distribution of stars and stellar remnants in an NSC of total mass $M_{\rm NSC}\sim 10^{6}M_{\odot}$, within the central ${\rm pc}^{3}$, then in the central $<10^{4}~R_{g}^{3}$ we expect O($10^{3}M_{\odot}$) drawn from the NSC population. From Fig~\ref{fig:stellar-Tcap_a} we expect around a quarter of these stars ($i<45^{\circ}$) to be captured on $\sim $Myr timescales into a SG-type disk, so within a factor of a few of the results in \citet{Bekdaulet2018}. Conversely, in a TQM-style AGN disk, we would only expect a small number of red giants within $<10^{4-5}~R_{g}$ to end up within the TQM disk, due to the longer capture times produced by the lower gas densities of the TQM disk. This illustrates the dependence of the disk population of stellar objects on gas disk properties, and highlights the importance of exploring the full parameter space of disk models.

\cite{MacLeod20} study the depletion of eccentric orbits into the AGN disk, allowing for interactions of nuclear orbiters with each other (via two-body relaxation) in addition to orbiter-disk interactions. They find disk capture has two phases. In the first phase, the orbital inclination, longitude of ascending node and argument of periapse remain nearly constant while ($a,e$) decrease rapidly, until the the orbits have circularized. In the second phase,  $i$ decreases until the orbiter has been captured by the disk.
\citet{MacLeod20}, used a disk model based on that of \citet{Rauch95}, who find that at $i \leq 30^{\circ}$, orbit circularization takes $\leq 0.1$Myr, which is less than the lowest value we assume for the fiducial AGN lifetime.  \citet{Rauch95} find the circularization time can become significant ($>0.1$Myr) at $i \geq 30^{\circ}$. By contrast with our findings, \citet{MacLeod20} find that sBH are unlikely to experience disk capture within an AGN disk lifetime.  However, (1) they utilize a very thin ($\alpha=1$) disk model \citep{Rauch95}, and so the fraction of an orbit experiencing drag force is very small and their $i_{crit}$ is larger; (2) sBH with initially highly eccentric orbits must first have their orbits circularized in order to be captured; but high eccentricity implies higher $v_{\rm rel}$ thereby yielding a larger $T_{\rm cap}$.  In our case, as shown in the top panels of Fig~\ref{fig:sBH-Tcap_a} and \ref{fig:sBH-Tcap_i} for the SG model, the sBH will be captured in $\leq 10$ Myrs for $i_{0} \leq 15^\circ$. This apparently small enhancement of the range of captured inclination angles significantly enhances the volume of the nucleus from which sBH can be captured by dense AGN disks. 

\subsection{Implications for LIGO-Virgo and LISA}
\label{sec:gw_imp}
\subsubsection{Dense AGN disks: $\rho >10^{-11}{\rm g \,cm^{-3}}$}
\label{sec:denseAGN}
 The capture of NSC orbits by AGN disks via drag can be relatively efficient for dense gas disks (e.g. SG-type models).  In this case, a large population of stars and sBH can be transferred from the NSC spheroid into the inner AGN disk over the disk lifetime. Once there, captured orbiters experience gas torques and migrate inward on a timescale inversely proportional to their mass (i.e. $\frac {da}{dt} \propto 1/M$, where $M$ is the migrator mass) \citep{Tanaka_2002}. The AGN disk models considered here possess migration traps where inward and outward gas torques cancel \citep{Bellovary16}. As a result, massive migrators (e.g. sBH) can encounter each other at small separations, form binaries and merge \citep{Secunda19,Secunda20}. Such sBH mergers are directly detectable with LIGO-Virgo and will tend to involve higher mass sBH as well as higher mass ratio mergers \citep{Yang19,McK19a}. A continuous supply of sBH to a disk migration trap as a result of disk capture would also support the build up of intermediate mass black holes (IMBH) at those migration traps \citep{McK19a}. The resulting population of IMBH-SMBH binaries should be detectable with LISA (particularly for SMBH $<10^{7}M_{\odot}$). 

Figure~\ref{fig:sBH-Tcap_a} shows that sBH on orbits with small inclination angles to the AGN disk will rapidly be captured. This result has implications for the re-capture of merged binary black holes.  Merging sBH binaries suffer recoil kicks due to GW emission \citep[e.g.][]{Campanelli07}, so merger remnants should end up on orbits with non-zero inclinations. However, the remnants will be re-captured by SG-type AGN disks quickly unless the post-kick inclination angle to the disk is $>15^{\circ}$. As an example, at $a=10^3 \, R_{g}$, where ${\rm v_{orb}}\sim10,000$km s$^{-1}$, a kick perpendicular to the disk midplane with ${\rm v_{kick}}<2500$ km s$^{-1}$ will still leave a remnant BH on an orbit with $i<15^{\circ}$. Thus, we expect AGN disks that are as dense as the SG model ($\rho \sim 10^{-11}{\rm g\,cm^{-3}}$) can retain or rapidly recapture the overwhelming majority of BBH merger products. Such AGN disks should thus be an ideal site for hierarchical BH merger scenarios. For lower density AGN disks, akin to TQM model disks ($\rho \sim 10^{-12}{\rm g\,cm^{-3}}$), a large fraction of kicked BBH mergers in the AGN channel will be lost from the disk. Thus, the relative rate of large mass BBH mergers as measured by LIGO-Virgo can help us probe typical properties of their host AGN disks, such as density and scale height as a function of disk radius.

\subsubsection{Low density AGN disks: $\rho <10^{-11}{\rm g \,cm^{-3}}$}
Drag is much less efficient for NSC orbit capture if AGN disks are generally more like the lower density TQM model. Several important implications follow. First, the population of embedded objects in those types of AGN disks will mostly correspond to those orbits initially geometrically coincident with the disk. This population will be a fraction O($H/a$) of the NSC population. Second, 
if AGN disks are typically low density, then the buildup of IMBH in this channel must be severely limited, even for relatively long lived AGN O(10s Myr). 

\subsection{Electromagnetic consequences}
\label{sec:EM}
Two types of EM counterparts should be expected from nuclear cluster objects interacting with AGN disks. First, embedded objects in AGN disks can generate  multiple EM signatures, due to interactions, accretion and merger of those embedded objects \citep[e.g.][]{McK20}. Second, objects that cross the disk but are not yet captured can yield EM flares due to disk entry and disk exit. In the latter case, the bow shock temperature from a disk-crossing star is $T_{\rm bow} \sim 10^{5}$K $({v}_{\rm rel}/10^{2}$ km s$^{-1})^{2}$ and the associated flare luminosity $L \sim 10^{38}$ erg s$^{-1} ({v}_{\rm rel}/10^{2}$ km s$^{-1})^{8} (R/R_{\odot})^{2}$ \citep{Zentsova83,McKernan14}. The $O$($v_{\rm rel}^{8}$) luminosity dependence of disk-crossing flares  suggests that higher inclination, large mass stars will produce very luminous, but short-lived flares from disk crossing. 

From \S\ref{sec:gw_imp}, dense AGN disks ($\rho \geq 10^{-11}{\rm g\, cm^{-3}}$) are likely to capture a significant fraction of nuclear cluster orbiters and to retain or re-capture kicked merger products, well within fiducial AGN lifetimes. As a result, EM counterparts due to an embedded population, such as supernovae, kilonovae and kicked BH merger products \citep{McK19b,Graham20b} will be more frequent in dense AGN disks. Conversely, and intruigingly, flaring due to disk crossing will be more common in lower density AGN disks; since capture is less efficient, a higher fraction of orbiters will exhibit plunging orbits. As a result, the observation that relatively high amplitude flaring is more common in low-luminosity AGN \citep{Hook94,Graham20} may be related to the lower efficiency of disk capture in low luminosity AGN, if many of those flares are due to disk crossing orbits. 

In addition, we can consider the energy deposited into the disk through the process of disk capture. A simple estimate of the change in potential energy from initial state to capture, divided by the capture time suggests a typical energy deposition $O(10^{38} \rm{erg/s})$ per object. While this is negligible compared to a bright AGN luminosity of $O(10^{45} \rm{erg/s})$, given many objects interacting with a disk (as in an NSC), the energy budget of the AGN may receive a substantial contribution from disk-crossers. This energy source may play a significant role in maintaining the outer disk against self-gravitating collapse (i.e. supporting Toomre's Q>1).

Even in the absence of direct EM counterparts to BBH mergers, \cite{Bartos17b} provides an elegant method for determining the AGN channel contribution to the LIGO-Virgo BBH merger detection rate. They compare the expected occurrence of AGN per LVC error volume to the observed occurrence of AGN per LVC error volume, and find that AGN-driven mergers should produce a small but measurable excess, accumulated over many detections. After accounting for realistic contamination rates in current galaxy catalogs and \citet{Ford19} find that after $\sim 600$ detected BBH mergers, we will be able to measure the AGN contribution to the merger rate, or limit it to $\leq 30\%$ of all such mergers at $\sim 95\%$ confidence. This measurement can help reveal the typical gas density of AGN disks, especially if that density is high, since high density disks capture more NSC objects and should produce more BBH mergers.

Separately, small scale mass loss due to repeated stellar encounters may contribute to the metallicity in AGN disks \citep{Artymowicz93}. Finally, embedded stars, or stars that spend some fraction of their orbits in the AGN disk may produce substantial metallicity enhancements due to their unusually large mass in- and outflow rates, caused by the altered boundary conditions experienced by stars embedded in AGN disks \citep{cantiello2020}. Such stars may also produce unique EM counterparts which may enable more direct probes of AGN disk densities and scale heights, among other properties.

\section{Conclusions}
\label{sec:conclusions}
We investigate the rate of capture of 
NSC orbiters by AGN disks around supermassive black holes. Our results depend strongly on the AGN disk model gas density. For the densest AGN disks ($\rho \geq 10^{-10}{\rm g \,cm^{-3}}$), much of the inner core of stars in the NSC can end up captured by the AGN disk by geometric drag forces for plausible estimates of AGN disk lifetimes. Stellar origin black holes (sBH) in NSCs at moderate inclination angles ($i<15^{\circ}$) are quickly captured by dense AGN disks, due to Bondi drag, and arrive in the inner disk. Thus, above a critical AGN disk density of $\rho \sim 10^{-11}{\rm g \,cm^{-3}}$ and disk lifetime $\geq 1$ Myr, we expect that there is a large embedded population of sBH, stars and stellar remnants which can support a significant population of mergers detectable with LIGO-Virgo and LISA. 

EM counterparts due to the embedded population will occur, as can signatures due to orbital capture. For less dense AGN disks ($\rho \leq 10^{-11}{\rm g \,cm^{-3}}$), drag is far less efficient and the embedded population will generally consist of orbits initially geometrically coincident with the AGN disk, with little support from captured orbits. However, lower density AGN disks should exhibit greater flaring variability due to disk crossing encounters. This latter point may account for some of the known anti-correlation of quasar luminosity with variability amplitude.

\section*{Acknowledgements}

GF \& SSN thank AstroCom NYC (NSF AST-1831415) for support, and Andreas Just and Bekdaulet Shukirgaliyev for useful insight on the dynamics of star-disk interactions. SSN thanks Mordecai-Mark Mac Low for computational resources, and Betsy Hernandez for peer mentorship. KESF \& BM are supported by NSF AST-1831415 and Simons Foundation Grant 533845. JMB is supported by NSF award AST-1812642. We thank the participants of the 1st Black Holes in the Disks of Active Galactic Nuclei Workshop held March 11-13, 2019 at the CCA, Flatiron Institute for very useful discussions. We would also like to thank the referee, Imre Bartos, for his helpful suggestions to elaborate on red giant-disk interactions, and to expand the discussion of the dependence of our results on AGN accretion rate; Cole Miller for his suggestion to compare energy deposition due to star-disk interactions to the disk luminosity; and Andrea Derdzinski for suggesting multiple edits which improved the clarity of the final work.

\section*{Data Availability}
The data underlying this article will be shared on request to the corresponding author.

\bibliographystyle{mnras}
\bibliography{grinddown} 

\begin{thebibliography}{}
\makeatletter
\relax
\def\mn@urlcharsother{\let\do\@makeother \do\$\do\&\do\#\do\^\do\_\do\%\do\~}
\def\mn@doi{\begingroup\mn@urlcharsother \@ifnextchar [ {\mn@doi@}
  {\mn@doi@[]}}
\def\mn@doi@[#1]#2{\def\@tempa{#1}\ifx\@tempa\@empty \href
  {http://dx.doi.org/#2} {doi:#2}\else \href {http://dx.doi.org/#2} {#1}\fi
  \endgroup}
\def\mn@eprint#1#2{\mn@eprint@#1:#2::\@nil}
\def\mn@eprint@arXiv#1{\href {http://arxiv.org/abs/#1} {{\tt arXiv:#1}}}
\def\mn@eprint@dblp#1{\href {http://dblp.uni-trier.de/rec/bibtex/#1.xml}
  {dblp:#1}}
\def\mn@eprint@#1:#2:#3:#4\@nil{\def\@tempa {#1}\def\@tempb {#2}\def\@tempc
  {#3}\ifx \@tempc \@empty \let \@tempc \@tempb \let \@tempb \@tempa \fi \ifx
  \@tempb \@empty \def\@tempb {arXiv}\fi \@ifundefined
  {mn@eprint@\@tempb}{\@tempb:\@tempc}{\expandafter \expandafter \csname
  mn@eprint@\@tempb\endcsname \expandafter{\@tempc}}}

\bibitem[\protect\citeauthoryear{{Alexander} \& {Hopman}}{{Alexander} \&
  {Hopman}}{2009}]{Alexander09}
{Alexander} T.,  {Hopman} C.,  2009, \mn@doi [\apj]
  {10.1088/0004-637X/697/2/1861}, \href
  {https://ui.adsabs.harvard.edu/abs/2009ApJ...697.1861A} {697, 1861}

\bibitem[\protect\citeauthoryear{{Antoni}, {MacLeod}  \&
  {Ramirez-Ruiz}}{{Antoni} et~al.}{2019}]{Antoni19}
{Antoni} A.,  {MacLeod} M.,   {Ramirez-Ruiz} E.,  2019, \mn@doi [\apj]
  {10.3847/1538-4357/ab3466}, 884

\bibitem[\protect\citeauthoryear{{Antonini}}{{Antonini}}{2014}]{Antonini14}
{Antonini} F.,  2014, \mn@doi [\apj] {10.1088/0004-637X/794/2/106}, \href
  {https://ui.adsabs.harvard.edu/abs/2014ApJ...794..106A} {794, 106}

\bibitem[\protect\citeauthoryear{{Armitage}, {Zurek}  \& {Davies}}{{Armitage}
  et~al.}{1996}]{Armitage96}
{Armitage} P.~J.,  {Zurek} W.~H.,   {Davies} M.~B.,  1996, \mn@doi [\apj]
  {10.1086/177864}, \href
  {https://ui.adsabs.harvard.edu/abs/1996ApJ...470..237A} {470, 237}

\bibitem[\protect\citeauthoryear{{Artymowicz} \& {Lubow}}{{Artymowicz} \&
  {Lubow}}{1994}]{ArtyLubow94}
{Artymowicz} P.,  {Lubow} S.~H.,  1994, \mn@doi [\apj] {10.1086/173679}, \href
  {https://ui.adsabs.harvard.edu/abs/1994ApJ...421..651A} {421, 651}

\bibitem[\protect\citeauthoryear{{Artymowicz}, {Lin}  \&
  {Wampler}}{{Artymowicz} et~al.}{1993}]{Artymowicz93}
{Artymowicz} P.,  {Lin} D.~N.~C.,   {Wampler} E.~J.,  1993, \mn@doi [\apj]
  {10.1086/172690}, \href
  {https://ui.adsabs.harvard.edu/abs/1993ApJ...409..592A} {409, 592}

\bibitem[\protect\citeauthoryear{{Bahcall} \& {Wolf}}{{Bahcall} \&
  {Wolf}}{1976}]{BahcallWolf76}
{Bahcall} J.~N.,  {Wolf} R.~A.,  1976, \mn@doi [\apj] {10.1086/154711}, \href
  {https://ui.adsabs.harvard.edu/abs/1976ApJ...209..214B} {209, 214}

\bibitem[\protect\citeauthoryear{{Bartos}, {Haiman}, {Marka}, {Metzger},
  {Stone}  \& {Marka}}{{Bartos} et~al.}{2017a}]{Bartos17b}
{Bartos} I.,  {Haiman} Z.,  {Marka} Z.,  {Metzger} B.~D.,  {Stone} N.~C.,
  {Marka} S.,  2017a, \mn@doi [Nature Communications]
  {10.1038/s41467-017-00851-7}, \href
  {https://ui.adsabs.harvard.edu/abs/2017NatCo...8..831B} {8, 831}

\bibitem[\protect\citeauthoryear{{Bartos}, {Kocsis}, {Haiman}  \&
  {M{\'a}rka}}{{Bartos} et~al.}{2017b}]{Bartos17}
{Bartos} I.,  {Kocsis} B.,  {Haiman} Z.,   {M{\'a}rka} S.,  2017b, \mn@doi
  [\apj] {10.3847/1538-4357/835/2/165}, \href
  {https://ui.adsabs.harvard.edu/abs/2017ApJ...835..165B} {835, 165}

\bibitem[\protect\citeauthoryear{{Baruteau}, {Cuadra}  \& {Lin}}{{Baruteau}
  et~al.}{2011}]{Baruteau11}
{Baruteau} C.,  {Cuadra} J.,   {Lin} D.~N.~C.,  2011, \mn@doi [\apj]
  {10.1088/0004-637X/726/1/28}, \href
  {https://ui.adsabs.harvard.edu/abs/2011ApJ...726...28B} {726, 28}

\bibitem[\protect\citeauthoryear{{Bellovary}, {Mac Low}, {McKernan}  \&
  {Ford}}{{Bellovary} et~al.}{2016}]{Bellovary16}
{Bellovary} J.~M.,  {Mac Low} M.-M.,  {McKernan} B.,   {Ford} K.~E.~S.,  2016,
  \mn@doi [\apj] {10.3847/2041-8205/819/2/L17}, \href
  {https://ui.adsabs.harvard.edu/abs/2016ApJ...819L..17B} {819, L17}

\bibitem[\protect\citeauthoryear{{B{\"o}ker}, {Laine}, {van der Marel},
  {Sarzi}, {Rix}, {Ho}  \& {Shields}}{{B{\"o}ker} et~al.}{2002}]{Boker02}
{B{\"o}ker} T.,  {Laine} S.,  {van der Marel} R.~P.,  {Sarzi} M.,  {Rix} H.-W.,
   {Ho} L.~C.,   {Shields} J.~C.,  2002, \mn@doi [\aj] {10.1086/339025}, \href
  {https://ui.adsabs.harvard.edu/abs/2002AJ....123.1389B} {123, 1389}

\bibitem[\protect\citeauthoryear{{Campanelli}, {Lousto}, {Zlochower}  \&
  {Merritt}}{{Campanelli} et~al.}{2007}]{Campanelli07}
{Campanelli} M.,  {Lousto} C.,  {Zlochower} Y.,   {Merritt} D.,  2007, \mn@doi
  [\apjl] {10.1086/516712}, \href
  {https://ui.adsabs.harvard.edu/abs/2007ApJ...659L...5C} {659, L5}

\bibitem[\protect\citeauthoryear{{Cannizzaro} et~al.,}{{Cannizzaro}
  et~al.}{2020}]{Cannizzaro20}
{Cannizzaro} G.,  et~al., 2020, \mn@doi [\mnras] {10.1093/mnras/staa186}, \href
  {https://ui.adsabs.harvard.edu/abs/2020MNRAS.tmp..180C} {p.~180}

\bibitem[\protect\citeauthoryear{Cantiello, Jermyn  \& Lin}{Cantiello
  et~al.}{2020}]{cantiello2020}
Cantiello M.,  Jermyn A.~S.,   Lin D. N.~C.,  2020, \href
  {https://ui.adsabs.harvard.edu/abs/2020arXiv200903936C} {p. arXiv:2009.03936}

\bibitem[\protect\citeauthoryear{{C{\^o}t{\'e}} et~al.,}{{C{\^o}t{\'e}}
  et~al.}{2006}]{Cote06}
{C{\^o}t{\'e}} P.,  et~al., 2006, \mn@doi [\apjs] {10.1086/504042}, \href
  {https://ui.adsabs.harvard.edu/abs/2006ApJS..165...57C} {165, 57}

\bibitem[\protect\citeauthoryear{{Derdzinski}, {D'Orazio}, {Duffell}, {Haiman}
  \& {MacFadyen}}{{Derdzinski} et~al.}{2019}]{Derdzinski19}
{Derdzinski} A.~M.,  {D'Orazio} D.,  {Duffell} P.,  {Haiman} Z.,   {MacFadyen}
  A.,  2019, \mn@doi [\mnras] {10.1093/mnras/stz1026}, \href
  {https://ui.adsabs.harvard.edu/abs/2019MNRAS.486.2754D} {486, 2754}

\bibitem[\protect\citeauthoryear{{Derdzinski}, {D'Orazio}, {Duffell}, {Haiman}
  \& {Macfadyen}}{{Derdzinski} et~al.}{2020}]{Derdzinski20}
{Derdzinski} A.,  {D'Orazio} D.,  {Duffell} P.,  {Haiman} Z.,   {Macfadyen} A.,
   2020, arXiv e-prints, \href
  {https://ui.adsabs.harvard.edu/abs/2020arXiv200511333D} {p. arXiv:2005.11333}

\bibitem[\protect\citeauthoryear{{Ford} et~al.,}{{Ford} et~al.}{2019}]{Ford19}
{Ford} K.~E.~S.,  et~al., 2019, \baas, \href
  {https://ui.adsabs.harvard.edu/abs/2019BAAS...51c.247F} {51, 247}

\bibitem[\protect\citeauthoryear{{Generozov}, {Stone}, {Metzger}  \&
  {Ostriker}}{{Generozov} et~al.}{2018}]{Generozov18}
{Generozov} A.,  {Stone} N.~C.,  {Metzger} B.~D.,   {Ostriker} J.~P.,  2018,
  \mn@doi [\mnras] {10.1093/mnras/sty1262}, \href
  {https://ui.adsabs.harvard.edu/abs/2018MNRAS.478.4030G} {478, 4030}

\bibitem[\protect\citeauthoryear{{Graham} \& {Spitler}}{{Graham} \&
  {Spitler}}{2009}]{Graham09}
{Graham} A.~W.,  {Spitler} L.~R.,  2009, \mn@doi [\mnras]
  {10.1111/j.1365-2966.2009.15118.x}, \href
  {https://ui.adsabs.harvard.edu/abs/2009MNRAS.397.2148G} {397, 2148}

\bibitem[\protect\citeauthoryear{{Graham}, {Djorgovski}, {Drake}, {Stern},
  {Mahabal}, {Glikman}, {Larson}  \& {Christensen}}{{Graham}
  et~al.}{2017}]{Graham17}
{Graham} M.~J.,  {Djorgovski} S.~G.,  {Drake} A.~J.,  {Stern} D.,  {Mahabal}
  A.~A.,  {Glikman} E.,  {Larson} S.,   {Christensen} E.,  2017, \mn@doi
  [\mnras] {10.1093/mnras/stx1456}, \href
  {https://ui.adsabs.harvard.edu/abs/2017MNRAS.470.4112G} {470, 4112}

\bibitem[\protect\citeauthoryear{{Graham} et~al.,}{{Graham}
  et~al.}{2020a}]{Graham20b}
{Graham} M.~J.,  et~al., 2020a, \mn@doi [\prl]
  {10.1103/PhysRevLett.124.251102}, \href
  {https://ui.adsabs.harvard.edu/abs/2020PhRvL.124y1102G} {124, 251102}

\bibitem[\protect\citeauthoryear{{Graham} et~al.,}{{Graham}
  et~al.}{2020b}]{Graham20}
{Graham} M.~J.,  et~al., 2020b, \mn@doi [\mnras] {10.1093/mnras/stz3244}, \href
  {https://ui.adsabs.harvard.edu/abs/2020MNRAS.491.4925G} {491, 4925}

\bibitem[\protect\citeauthoryear{{Gr{\"o}bner}, {Ishibashi}, {Tiwari}, {Haney}
  \& {Jetzer}}{{Gr{\"o}bner} et~al.}{2020}]{Grobner20}
{Gr{\"o}bner} M.,  {Ishibashi} W.,  {Tiwari} S.,  {Haney} M.,   {Jetzer} P.,
  2020, \mn@doi [A&A] {10.1051/0004-6361/202037681}, \href
  {https://ui.adsabs.harvard.edu/abs/2020arXiv200503571G} {638, 8}

\bibitem[\protect\citeauthoryear{{Haehnelt} \& {Rees}}{{Haehnelt} \&
  {Rees}}{1993}]{Haehnelt93}
{Haehnelt} M.~G.,  {Rees} M.~J.,  1993, \mn@doi [\mnras]
  {10.1093/mnras/263.1.168}, \href
  {https://ui.adsabs.harvard.edu/abs/1993MNRAS.263..168H} {263, 168}

\bibitem[\protect\citeauthoryear{{Hailey}, {Mori}, {Bauer}, {Berkowitz}, {Hong}
   \& {Hord}}{{Hailey} et~al.}{2018}]{Hailey18}
{Hailey} C.~J.,  {Mori} K.,  {Bauer} F.~E.,  {Berkowitz} M.~E.,  {Hong} J.,
  {Hord} B.~J.,  2018, \mn@doi [\nat] {10.1038/nature25029}, \href
  {https://ui.adsabs.harvard.edu/abs/2018Natur.556...70H} {556, 70}

\bibitem[\protect\citeauthoryear{{Hook}, {McMahon}, {Boyle}  \& {Irwin}}{{Hook}
  et~al.}{1994}]{Hook94}
{Hook} I.~M.,  {McMahon} R.~G.,  {Boyle} B.~J.,   {Irwin} M.~J.,  1994, \mn@doi
  [\mnras] {10.1093/mnras/268.2.305}, \href
  {https://ui.adsabs.harvard.edu/abs/1994MNRAS.268..305H} {268, 305}

\bibitem[\protect\citeauthoryear{Horn, Lyra, Low  \& S{\'{a}}ndor}{Horn
  et~al.}{2012}]{Horn_2012}
Horn B.,  Lyra W.,  Low M.-M.~M.,   S{\'{a}}ndor Z.,  2012, \mn@doi [ApJ]
  {10.1088/0004-637x/750/1/34}, 750, 34

\bibitem[\protect\citeauthoryear{{Ishibashi} \& {Gr{\"o}bner}}{{Ishibashi} \&
  {Gr{\"o}bner}}{2020}]{Ishibashi20}
{Ishibashi} W.,  {Gr{\"o}bner} M.,  2020, \mn@doi [A&A]
  {10.1051/0004-6361/202037799}, \href
  {https://ui.adsabs.harvard.edu/abs/2020arXiv200607407I} {639, 9}

\bibitem[\protect\citeauthoryear{{Just}, {Yurin}, {Makukov}, {Berczik},
  {Chingis}, {Spurzem}  \& {Vilkoviski}}{{Just} et~al.}{2012}]{Just12}
{Just} A.,  {Yurin} D.,  {Makukov} {Berczik} P.,  {Chingis} O.,  {Spurzem} R.,
   {Vilkoviski} E.~Y.,  2012, \mn@doi [\apj] {10.1088/0004-637x/758/1/51}, 758,
  51

\bibitem[\protect\citeauthoryear{{Kennedy}, {Meiron}, {Shukirgaliyev},
  {Panamarev}, {Berczik}, {Just}  \& {Spurzem}}{{Kennedy}
  et~al.}{2016}]{Kennedy16}
{Kennedy} G.~F.,  {Meiron} Y.,  {Shukirgaliyev} B.,  {Panamarev} T.,  {Berczik}
  P.,  {Just} A.,   {Spurzem} R.,  2016, \mn@doi [\mnras]
  {10.1093/mnras/stw908}, \href
  {https://ui.adsabs.harvard.edu/abs/2016MNRAS.460..240K} {460, 240}

\bibitem[\protect\citeauthoryear{{Kieffer} \& {Bogdanovi{\'c}}}{{Kieffer} \&
  {Bogdanovi{\'c}}}{2016}]{Kieffer16}
{Kieffer} T.~F.,  {Bogdanovi{\'c}} T.,  2016, \mn@doi [\apj]
  {10.3847/0004-637X/823/2/155}, \href
  {https://ui.adsabs.harvard.edu/abs/2016ApJ...823..155K} {823, 155}

\bibitem[\protect\citeauthoryear{{King} \& {Nixon}}{{King} \&
  {Nixon}}{2015}]{King15}
{King} A.,  {Nixon} C.,  2015, \mn@doi [\mnras] {10.1093/mnrasl/slv098}, \href
  {https://ui.adsabs.harvard.edu/abs/2015MNRAS.453L..46K} {453, L46}

\bibitem[\protect\citeauthoryear{{Leigh}, {B{\"o}ker}  \& {Knigge}}{{Leigh}
  et~al.}{2012}]{Leigh12}
{Leigh} N.,  {B{\"o}ker} T.,   {Knigge} C.,  2012, \mn@doi [\mnras]
  {10.1111/j.1365-2966.2012.21365.x}, \href
  {https://ui.adsabs.harvard.edu/abs/2012MNRAS.424.2130L} {424, 2130}

\bibitem[\protect\citeauthoryear{{Leigh} et~al.,}{{Leigh}
  et~al.}{2018}]{Leigh18}
{Leigh} N.~W.~C.,  et~al., 2018, \mn@doi [\mnras] {10.1093/mnras/stx3134},
  \href {https://ui.adsabs.harvard.edu/abs/2018MNRAS.474.5672L} {474, 5672}

\bibitem[\protect\citeauthoryear{Levin}{Levin}{2007}]{Levin_2007}
Levin Y.,  2007, \mn@doi [MNRAS] {10.1111/j.1365-2966.2006.11155.x}, 374, 515

\bibitem[\protect\citeauthoryear{{Lin} \& {Papaloizou}}{{Lin} \&
  {Papaloizou}}{1986}]{LinPap86}
{Lin} D.~N.~C.,  {Papaloizou} J.,  1986, \mn@doi [\apj] {10.1086/164653}, \href
  {https://ui.adsabs.harvard.edu/abs/1986ApJ...309..846L} {309, 846}

\bibitem[\protect\citeauthoryear{{Lubow}}{{Lubow}}{1981}]{Lubow81}
{Lubow} S.~H.,  1981, \mn@doi [\apj] {10.1086/158808}, \href
  {https://ui.adsabs.harvard.edu/abs/1981ApJ...245..274L} {245, 274}

\bibitem[\protect\citeauthoryear{Lyra, Paardekooper  \& Low}{Lyra
  et~al.}{2010}]{Lyra_2010}
Lyra W.,  Paardekooper S.-J.,   Low M.-M.~M.,  2010, \mn@doi [ApJ]
  {10.1088/2041-8205/715/2/l68}, 715, L68

\bibitem[\protect\citeauthoryear{{MacLeod} \& {Lin}}{{MacLeod} \&
  {Lin}}{2020}]{MacLeod20}
{MacLeod} M.,  {Lin} D. N.~C.,  2020, \mn@doi [\apj]
  {10.3847/1538-4357/ab64db}, \href
  {https://ui.adsabs.harvard.edu/abs/2020ApJ...889...94M} {889, 94}

\bibitem[\protect\citeauthoryear{McKernan, Ford, Lyra  \& Perets}{McKernan
  et~al.}{2012}]{McKernan12}
McKernan B.,  Ford K. E.~S.,  Lyra W.,   Perets H.~B.,  2012, \mn@doi [MNRAS]
  {10.1111/j.1365-2966.2012.21486.x}, 425, 460

\bibitem[\protect\citeauthoryear{McKernan, Ford, Kocsis, Lyra  \&
  Winter}{McKernan et~al.}{2014}]{McKernan14}
McKernan B.,  Ford K. E.~S.,  Kocsis B.,  Lyra W.,   Winter L.~M.,  2014,
  \mn@doi [MNRAS] {10.1093/mnras/stu553}, 441, 900

\bibitem[\protect\citeauthoryear{{McKernan} et~al.,}{{McKernan}
  et~al.}{2018}]{McKernan18}
{McKernan} B.,  et~al., 2018, \mn@doi [\apj] {10.3847/1538-4357/aadae5}, \href
  {https://ui.adsabs.harvard.edu/abs/2018ApJ...866...66M} {866, 66}

\bibitem[\protect\citeauthoryear{{McKernan} et~al.,}{{McKernan}
  et~al.}{2019}]{McK19b}
{McKernan} B.,  et~al., 2019, \mn@doi [\apjl] {10.3847/2041-8213/ab4886}, \href
  {https://ui.adsabs.harvard.edu/abs/2019ApJ...884L..50M} {884, L50}

\bibitem[\protect\citeauthoryear{{McKernan}, {Ford}, {O'Shaugnessy}  \&
  {Wysocki}}{{McKernan} et~al.}{2020a}]{McK19a}
{McKernan} B.,  {Ford} K.~E.~S.,  {O'Shaugnessy} R.,   {Wysocki} D.,  2020a,
  \mn@doi [\mnras] {10.1093/mnras/staa740}, \href
  {https://ui.adsabs.harvard.edu/abs/2020MNRAS.494.1203M} {494, 1203}

\bibitem[\protect\citeauthoryear{{McKernan}, {Ford}  \&
  {O'Shaughnessy}}{{McKernan} et~al.}{2020b}]{McK20}
{McKernan} B.,  {Ford} K.~E.~S.,   {O'Shaughnessy} R.,  2020b, \mn@doi [\mnras]
  {10.1093/mnras/staa2681}, \href
  {https://ui.adsabs.harvard.edu/abs/2020arXiv200200046M} {498, 4088}

\bibitem[\protect\citeauthoryear{{McLaughlin}, {King}  \&
  {Nayakshin}}{{McLaughlin} et~al.}{2006}]{McLaughlin06}
{McLaughlin} D.~E.,  {King} A.~R.,   {Nayakshin} S.,  2006, \mn@doi [\apjl]
  {10.1086/508627}, \href
  {https://ui.adsabs.harvard.edu/abs/2006ApJ...650L..37M} {650, L37}

\bibitem[\protect\citeauthoryear{{Miralda-Escud{\'e}} \&
  {Gould}}{{Miralda-Escud{\'e}} \& {Gould}}{2000}]{MiraldaEscude00}
{Miralda-Escud{\'e}} J.,  {Gould} A.,  2000, \mn@doi [\apj] {10.1086/317837},
  \href {https://ui.adsabs.harvard.edu/abs/2000ApJ...545..847M} {545, 847}

\bibitem[\protect\citeauthoryear{{Morris}}{{Morris}}{1993}]{Morris93}
{Morris} M.,  1993, \mn@doi [\apj] {10.1086/172607}, \href
  {https://ui.adsabs.harvard.edu/abs/1993ApJ...408..496M} {408, 496}

\bibitem[\protect\citeauthoryear{{Netzer}}{{Netzer}}{2015}]{Netzer15}
{Netzer} H.,  2015, \mn@doi [\araa] {10.1146/annurev-astro-082214-122302},
  \href {https://ui.adsabs.harvard.edu/abs/2015ARA&A..53..365N} {53, 365}

\bibitem[\protect\citeauthoryear{{Neumayer}, {Seth}  \& {Boeker}}{{Neumayer}
  et~al.}{2020}]{Neumayer20}
{Neumayer} N.,  {Seth} A.,   {Boeker} T.,  2020, \mn@doi [A&A]
  {10.1007/s00159-020-00125-0}, \href
  {https://ui.adsabs.harvard.edu/abs/2020arXiv200103626N} {28, 4}

\bibitem[\protect\citeauthoryear{{Novikov} \& {Thorne}}{{Novikov} \&
  {Thorne}}{1973}]{NovikovThorne73}
{Novikov} I.~D.,  {Thorne} K.~S.,  1973, in Black Holes (Les Astres Occlus). pp
  343--450

\bibitem[\protect\citeauthoryear{{Ostriker}}{{Ostriker}}{1983}]{Ostriker83}
{Ostriker} J.~P.,  1983, \mn@doi [\apj] {10.1086/161351}, \href
  {https://ui.adsabs.harvard.edu/abs/1983ApJ...273...99O} {273, 99}

\bibitem[\protect\citeauthoryear{{Ostriker}}{{Ostriker}}{1999}]{Ostriker99}
{Ostriker} E.~C.,  1999, \mn@doi [\apj] {10.1086/306858}, \href
  {https://ui.adsabs.harvard.edu/abs/1999ApJ...513..252O} {513, 252}

\bibitem[\protect\citeauthoryear{Paardekooper, Baruteau, Crida  \&
  Kley}{Paardekooper et~al.}{2010}]{Paardekooper_2010}
Paardekooper S.-J.,  Baruteau C.,  Crida A.,   Kley W.,  2010, \mn@doi [MNRAS]
  {10.1111/j.1365-2966.2009.15782.x}, 401, 1950

\bibitem[\protect\citeauthoryear{{Panamarev}, {Shukirgaliyev}  \&
  {Meiron}}{{Panamarev} et~al.}{2018}]{Bekdaulet2018}
{Panamarev} T.,  {Shukirgaliyev} B.,   {Meiron} Y.,  2018, \mn@doi [\mnras]
  {10.1093/mnras/sty459}, 476, 4224

\bibitem[\protect\citeauthoryear{{Passy}, {Mac Low}  \& {De Marco}}{{Passy}
  et~al.}{2012}]{Passy12}
{Passy} J.-C.,  {Mac Low} M. M.~K.,   {De Marco} O.,  2012, \mn@doi [\apj]
  {10.1088/2041-8205/759/2/L30}, \href
  {https://ui.adsabs.harvard.edu/abs/2012ApJ...759L..30P/abstract} {866}

\bibitem[\protect\citeauthoryear{{Preto} \& {Amaro-Seoane}}{{Preto} \&
  {Amaro-Seoane}}{2010}]{PretoAmaro10}
{Preto} M.,  {Amaro-Seoane} P.,  2010, \mn@doi [\apjl]
  {10.1088/2041-8205/708/1/L42}, \href
  {https://ui.adsabs.harvard.edu/abs/2010ApJ...708L..42P} {708, L42}

\bibitem[\protect\citeauthoryear{{Rauch}}{{Rauch}}{1995}]{Rauch95}
{Rauch} K.~P.,  1995, \mn@doi [\mnras] {10.1093/mnras/275.3.628}, 275, 628

\bibitem[\protect\citeauthoryear{{Schawinski}, {Koss}, {Berney}  \&
  {Sartori}}{{Schawinski} et~al.}{2015}]{Schawinski15}
{Schawinski} K.,  {Koss} M.,  {Berney} S.,   {Sartori} L.~F.,  2015, \mn@doi
  [\mnras] {10.1093/mnras/stv1136}, \href
  {https://ui.adsabs.harvard.edu/abs/2015MNRAS.451.2517S} {451, 2517}

\bibitem[\protect\citeauthoryear{{Scott} \& {Graham}}{{Scott} \&
  {Graham}}{2013}]{Scott13}
{Scott} N.,  {Graham} A.~W.,  2013, \mn@doi [\apj]
  {10.1088/0004-637X/763/2/76}, \href
  {https://ui.adsabs.harvard.edu/abs/2013ApJ...763...76S} {763, 76}

\bibitem[\protect\citeauthoryear{{Secunda}, {Bellovary}, {Mac Low}, {Ford},
  {McKernan}, {Leigh}, {Lyra}  \& {S{\'a}ndor}}{{Secunda}
  et~al.}{2019}]{Secunda19}
{Secunda} A.,  {Bellovary} J.,  {Mac Low} M.-M.,  {Ford} K.~E.~S.,  {McKernan}
  B.,  {Leigh} N. W.~C.,  {Lyra} W.,   {S{\'a}ndor} Z.,  2019, \mn@doi [\apj]
  {10.3847/1538-4357/ab20ca}, \href
  {https://ui.adsabs.harvard.edu/abs/2019ApJ...878...85S} {878, 85}

\bibitem[\protect\citeauthoryear{{Secunda} et~al.,}{{Secunda}
  et~al.}{2020}]{Secunda20}
{Secunda} A.,  et~al., 2020, arXiv e-prints, \href
  {https://ui.adsabs.harvard.edu/abs/2020arXiv200411936S} {p. arXiv:2004.11936}

\bibitem[\protect\citeauthoryear{{Shakura} \& {Sunyaev}}{{Shakura} \&
  {Sunyaev}}{1973}]{SS73}
{Shakura} N.~I.,  {Sunyaev} R.~A.,  1973, \aap, \href
  {https://ui.adsabs.harvard.edu/abs/1973A&A....24..337S} {500, 33}

\bibitem[\protect\citeauthoryear{Sirko \& Goodman}{Sirko \& Goodman}{2003}]{SG}
Sirko E.,  Goodman J.,  2003, \mn@doi [MNRAS]
  {10.1046/j.1365-8711.2003.06431.x}, 341, 501

\bibitem[\protect\citeauthoryear{{Stone}, {Metzger}  \& {Haiman}}{{Stone}
  et~al.}{2017}]{Stone17}
{Stone} N.~C.,  {Metzger} B.~D.,   {Haiman} Z.,  2017, \mn@doi [\mnras]
  {10.1093/mnras/stw2260}, \href
  {https://ui.adsabs.harvard.edu/abs/2017MNRAS.464..946S} {464, 946}

\bibitem[\protect\citeauthoryear{{Syer}, {Clarke}  \& {Rees}}{{Syer}
  et~al.}{1991}]{Syer91}
{Syer} D.,  {Clarke} C.~J.,   {Rees} M.~J.,  1991, \mn@doi [\mnras]
  {10.1093/mnras/250.3.505}, \href
  {https://ui.adsabs.harvard.edu/abs/1991MNRAS.250..505S} {250, 505}

\bibitem[\protect\citeauthoryear{{Tagawa}, {Haiman}, {Bartos}  \&
  {Kocsis}}{{Tagawa} et~al.}{2020}]{Tagawa20}
{Tagawa} H.,  {Haiman} Z.,  {Bartos} I.,   {Kocsis} B.,  2020, \mn@doi [\apj]
  {doi.org/10.3847/1538-4357/aba2cc}, \href
  {https://ui.adsabs.harvard.edu/abs/2020arXiv200411914T} {899, 26}

\bibitem[\protect\citeauthoryear{Tanaka, Takeuchi  \& Ward}{Tanaka
  et~al.}{2002}]{Tanaka_2002}
Tanaka H.,  Takeuchi T.,   Ward W.~R.,  2002, \mn@doi [ApJ] {10.1086/324713},
  565, 1257

\bibitem[\protect\citeauthoryear{{The LIGO Scientific Collaboration} \& {the
  Virgo Collaboration}}{{The LIGO Scientific Collaboration} \& {the Virgo
  Collaboration}}{2020a}]{GWa}
{The LIGO Scientific Collaboration} {the Virgo Collaboration} 2020a, \mn@doi
  [Phys. Rev. Lett.] {10.1103/PhysRevLett.125.101102}, \href
  {https://ui.adsabs.harvard.edu/abs/2020arXiv200901075T} {125, 101102}

\bibitem[\protect\citeauthoryear{{The LIGO Scientific Collaboration} \& {the
  Virgo Collaboration}}{{The LIGO Scientific Collaboration} \& {the Virgo
  Collaboration}}{2020b}]{GWb}
{The LIGO Scientific Collaboration} {the Virgo Collaboration} 2020b, \mn@doi
  [Astrophys. J. Lett.] {10.3847/2041-8213/aba493}, 900, L13

\bibitem[\protect\citeauthoryear{Thompson, Quataert  \& Murray}{Thompson
  et~al.}{2005}]{TQM}
Thompson T.~A.,  Quataert E.,   Murray N.,  2005, \mn@doi [ApJ]
  {10.1086/431923}, 630, 167

\bibitem[\protect\citeauthoryear{Ward}{Ward}{1997}]{Ward_1997}
Ward W.,  1997, \mn@doi [Icarus] {10.1006/icar.1996.5647}, 126, 261

\bibitem[\protect\citeauthoryear{{Wehner} \& {Harris}}{{Wehner} \&
  {Harris}}{2006}]{Wehner06}
{Wehner} E.~H.,  {Harris} W.~E.,  2006, \mn@doi [\apjl] {10.1086/505387}, \href
  {https://ui.adsabs.harvard.edu/abs/2006ApJ...644L..17W} {644, L17}

\bibitem[\protect\citeauthoryear{{Yang} et~al.,}{{Yang} et~al.}{2019}]{Yang19}
{Yang} Y.,  et~al., 2019, \mn@doi [\prl] {10.1103/PhysRevLett.123.181101},
  \href {https://ui.adsabs.harvard.edu/abs/2019PhRvL.123r1101Y} {123, 181101}

\bibitem[\protect\citeauthoryear{{Zentsova}}{{Zentsova}}{1983}]{Zentsova83}
{Zentsova} A.~S.,  1983, \mn@doi [\apss] {10.1007/BF00661152}, \href
  {https://ui.adsabs.harvard.edu/abs/1983Ap&SS..95...11Z} {95, 11}

\makeatother
\end{thebibliography}

\bsp
\label{lastpage}
\end{document}